\documentclass[11pt,]{article}
\usepackage{authblk}
\usepackage{fullpage}
\usepackage{amssymb,amsmath}
\usepackage[utf8x]{inputenc}
\usepackage[T1]{fontenc}
\usepackage{siunitx}
\usepackage[version=3]{mhchem}
\usepackage{xr}
\usepackage{bm}
\usepackage{pdfpages}
\usepackage{graphicx}% Include figure files
\usepackage{dcolumn}% Align table columns on decimal point
\usepackage{bm}% bold math
\usepackage{color}
\usepackage{tikz}
\usepackage{setspace}
\usepackage[title]{appendix}
\newcommand*\emptycirc[1][0.9ex]{\tikz\draw (0,0) circle (#1);}
\usepackage{float}
\usepackage{braket}
\usepackage[center,font=small,labelfont=bf]{caption}

\usepackage[square,numbers,sort&compress]{natbib}
\usepackage[left]{lineno}
\linenumbers

\usepackage{setspace}
\usepackage[unicode=true]{hyperref}
\hypersetup{colorlinks=true,
            citecolor=blue,
            linkcolor=blue}
\newcommand{\s}[1]{\mbox{\tiny{#1}}}
\numberwithin{equation}{section}

\usepackage{graphicx}
\graphicspath{{figures/}}
\makeatletter
\def\ScaleWidthIfNeeded{%
 \ifdim\Gin@nat@width>\linewidth
    \linewidth
  \else
    \Gin@nat@width
  \fi
}
\def\ScaleHeightIfNeeded{%
  \ifdim\Gin@nat@height>0.9\textheight
    0.9\textheight
  \else
    \Gin@nat@width
  \fi
}
\makeatother
\setkeys{Gin}{width=\ScaleWidthIfNeeded,height=\ScaleHeightIfNeeded,keepaspectratio}%

\newcommand{\corr}[1]{{\color{red}{#1}}}

\begin{document}
\nolinenumbers

\title{Demographic effects of aggregation in the presence of a component Allee effect}
%\title{Emergence of spatial patterns and Allee effect in a population with nonlocal competition and facilitation}
%\title{Population dynamics with nonlocal competition and facilitation: spatial patterns and Allee effect.}
%1-Survival and extinction in a non-local model for Allee effect\\2- Positive and negative interactions with different scales driving pattern formation in populations\\3- \ricardo{Non-local scale-dependent feedbacks in systems of interacting particles}\\
%4. \ricardo{Allee effect and competing scales}\\(Suggestions for titles)}% Force line breaks with \\
%\thanks{A footnote to the article title}

\author[1,2]{Daniel C.P. Jorge}
\author[3,1*]{Ricardo Martinez-Garcia}
\affil[1]{\footnotesize ICTP South American Institute for Fundamental Research \& Instituto de F\'isica Te\'orica, Universidade Estadual Paulista - UNESP, Rua Dr. Bento Teobaldo Ferraz 271, Bloco 2 - Barra Funda, 01140-070 S\~ao Paulo, SP, Brazil}
\affil[2]{Department of Ecology and Evolutionary Biology, Princeton University, Princeton NJ 08544, USA.}
\affil[3]{Center for Advanced Systems Understanding (CASUS); Helmholtz-Zentrum Dresden Rossendorf (HZDR), G{\"o}rlitz, Germany.}
 \affil[*]{Corresponding author: r.martinez-garcia@hzdr.de}

\date{}% It is always \today, today,

\maketitle 

%\newpag

\begin{abstract}
The component Allee effect (AE) is the positive correlation between an organism’s fitness component and population density. Depending on the population spatial structure, which determines the interactions between organisms, a component AE might lead to positive density-dependence in the population per capita growth rate and establish a demographic AE. However, existing spatial models impose a fixed population spatial structure, which limits the understanding of how a component AE and spatial dynamics jointly determine the existence of demographic AEs. We introduce a spatially explicit theoretical framework where spatial structure and population dynamics are emergent properties of the individual-level demographic and movement rates. This framework predicts various spatial patterns depending on its specific parameterization, including evenly spaced aggregates of organisms, that determine the demographic-level by-products of the component AE. We find that aggregation increases population abundance and allows population survival in harsher environments and at lower global population densities when compared with uniformly distributed organisms. Moreover, aggregation can prevent the component AE from manifesting at the population level or restrict it to the level of each independent aggregate. These results provide a mechanistic understanding of how component AEs might operate for different spatial structures and manifest at larger scales.
\end{abstract}
%\newpage

%\tableofcontents

\section{\label{Intro}Introduction}

Intraspecific interactions are key to understanding population ecology because they define how demographic rates depend on population density and ultimately drive population dynamics. The Allee effect is characterized by a positive correlation between population size or density and any individual fitness component \citep{stephens1999allee,levitan2005,courchamp2008allee}. Because of this positive density dependence, populations subjected to Allee effects may exhibit thresholds for population survival that manifest in sudden extinctions, \corr{the} existence of alternative stable states, and hysteresis loops \citep{lande1987extinction,courchamp2008allee,sun2016mathematical,oro2020ch6}. These highly nonlinear features make populations exhibiting Allee effects hard to manage without a mechanistic understanding of how individual-level processes and interactions lead to the trends and patterns observed in population dynamics.

\begin{figure}[ht]\centering\centering
\includegraphics[width={0.7\linewidth}]{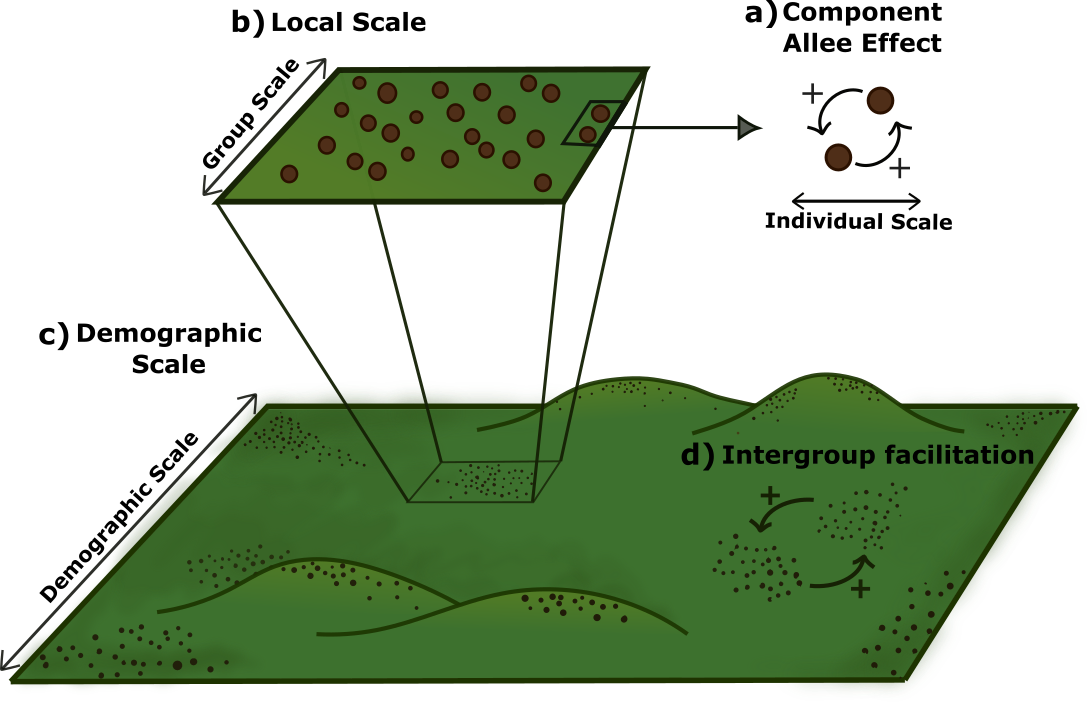}
\caption{\label{fig:0} Allee effect across spatial scales. The component Allee effect (a) results from interactions between individuals, and it manifests at a (b) local scale around a focal organism. Individuals are spatially scattered at the demographic scale (c), possibly forming aggregates. In the presence of aggregates, the population has a fourth characteristic scale, defining inter-group facilitation (d)}
\end{figure}

Allee effects are studied mainly at two levels: the component and the demographic Allee effect \citep{stephens1999allee}. The component Allee effect is a positive association between population density and one (or many \cite{berec2007}) individual fitness components, such as offspring survival, mating success, or fecundity \citep{orr2009fitness,drake2011allee, courchamp2008allee} (Fig.\,\ref{fig:0}a). Component Allee effects rely on several mechanisms. In some fish, rotifer, and mammals such as marmots, the presence of conspecifics changes the environmental conditions locally, improving habitat quality and individual fitness \citep{allee1932studies, stephens2002model,allee1949group,ghazoul2005buzziness}. Especially in group-living organisms, cooperative behaviors such as group vigilance, nursing, resource sharing, and social foraging also make individuals more competent in the presence of conspecifics \citep{dechmann2010group, nowak2011demographic,snaith2008red,angulo2013social,luque2013allee, angulo2018allee}. Allee effects are also frequent in sexually reproducing species. In motile organisms, females are more likely to find mates at larger population sizes \citep{dennis1989allee, liermann2001depensation,tcheslavskaia2002mating,garrett2002allee}. In sessile organisms, such as pollinators or broadcast spawners, fecundation is more likely at high population densities \citep{ashman2004pollen,wagenius2006scale,luzuriaga2006population, lundquist2011estimating, guy2019importance}. On the other hand, the demographic Allee effect is a population-level feature that emerges from one or more component Allee effects and manifests as a positive correlation between the net per-capita growth rate and the population size. This positive density-dependence is easier to identify at low population densities because competition hinders its effect in more crowded scenarios \citep{Morris2002,courchamp2008allee}. Demographic Allee effects are strong if the population cannot survive below a specific threshold size (Allee threshold) or weak if positive density-dependence is not intense enough to establish such a survival threshold \citep{courchamp2008allee,drake2011allee}.

The fitness of a focal individual in the presence of a component Allee effect is a nonlinear function of the local density of conspecifics around it (Fig.\,\ref{fig:0}b). Moreover, because Allee effects have a stronger impact at low population densities and often require the direct interaction between at least two organisms, the spatial population structure determines whether and to which extent a component Allee effect will lead to a demographic one \citep{Cassini2011,kramer2009evidence, kanarek2013allee,Surendran2020} (Fig.\,\ref{fig:0}c). Back to Allee's seminal experiments, several studies have investigated the impact of the spatial population structure, and more specifically of aggregation, on Allee effects \citep{allee1938social}. For example, some plant populations produce more and heavier seeds if distributed in clumps \citep{luzuriaga2006population,wagenius2006scale}. Plant aggregates can also facilitate nearby individuals because they attract pollinators to them, which extends the facilitation range beyond the scale of a single cluster of plants (Fig.\,\ref{fig:0}d) \citep{le2008allee}, and ameliorate physical stresses \citep{silliman2015facilitation}. Broadcast spawners subjected to a strong Allee effect, such as the red sea urchin \textit{Strongylocentrotus franciscanus}, can survive at low abundances by aggregating \citep{lundquist2011estimating,guy2019importance}. Finally, several social species form spatially segregated groups, which could contribute to population persistence in harsh environmental conditions \citep{angulo2018allee,lerch2018demographic, woodroffe2020within}. Aggregation and group living are thus ubiquitous features of populations subjected to Allee effects, and they strongly influence the emergent population dynamics. To explain how these spatial features impact populations subjected to component Allee effects, recent studies have introduced the group-level Allee effect, defined as any positive association between the organism's fitness and group size \citep{lerch2018demographic}. However, a theoretical framework that simultaneously describes how group-level Allee effects emerge from component Allee effects and the individual-level processes responsible for aggregation and group formation is lacking.

Much of the current understanding of Allee effects has developed from theoretical studies \citep{volterra1938, kostitzin1940,tammes1964sexual,hsu1975population,asmussen1979density,lande1987extinction,cushing1988,sun2016mathematical}. Several models, either deterministic or stochastic, consider well-mixed populations and disregard spatial degrees of freedom \citep{dennis1981extinction,dennis2002,mendez2019demographic}. The effect of space has been investigated mainly using metapopulation approaches in which each node represents a group or cluster of individuals and links represent interactions between groups \citep{rijnsdorp2001feeding,padron2000effect}. These frameworks already incorporate group-level Allee effects because they restrict fitness benefits due to intraspecific interactions to each metapopulation, and have helped explain why component Allee effects rarely manifest at the demographic level in group-living species \citep{rijnsdorp2001feeding,courchamp2008allee}. However, metapopulation models impose the existence of groups in the stationary state and do not describe the group-forming dynamics. Alternative approaches, based on individual-based models (IBMs) or partial differential equations (PDEs), incorporate space explicitly and can describe the group-forming dynamics \citep{Fadai2019,keitt2001allee,maciel2015allee,Surendran2020,takasu2009,berec2001,boukal2002}. Therefore, these approaches can explain how different spatial patterns of population density impact the outcome of ecological dynamics, such as species invasions \citep{maciel2015allee,keitt2001allee}, in the presence of Allee effects or Allee-effect features, such as the Allee threshold \citep{Surendran2020} and how it depends on the spatial scales characteristic of the individual-level interactions \citep{boukal2002}.

In this work, we develop a theoretical framework to investigate Allee effects across different levels of spatial organization within a population (Fig.\,\ref{fig:0}). We present this formalism starting from a stochastic and spatially explicit individual-based description of a population with density-dependent reproduction mimicking a component Allee effect. This description is the most fundamental level at which we can describe a population, allowing us to explicitly model the relationship between the mechanism responsible for the component Allee effect and individual birth and death rates. From this individual-level description, we derive the corresponding deterministic PDE for the dynamics of the population density. This approximation allows us to investigate the conditions in which individuals aggregate due to individual-level interactions and to study the population-level consequences of the component Allee effect depending on the spatial population structure. Finally, we identify the cases in which we can describe the long-term spatial distribution of individuals in terms of a metapopulation model and use this approach to investigate the emergence of group-level Allee effects. Our results recapitulate several empirical observations regarding the influence of spatial structure on group and demographic Allee effects, providing a unifying theoretical framework to investigate the interplay between component Allee effects and spatial dynamics. %Moreover, because population-level dynamics and patterns emerge from individual-level processes in our model, we can understand the origin of these observed behaviors at the most fundamental level and frame them within a common theoretical framework.

\section{\label{sec1} Methods}
\subsection{A spatially explicit individual-based model with component Allee effect}
\label{sec1:level1}

At the most fundamental level, we describe the spatiotemporal population dynamics using an individual-based model (IBM) in which we can incorporate any ecological interaction, such as competition, predation, or cooperation, movement, and birth-death dynamics at the organism level \citep{mendez2019demographic,hernandez2004clustering, Surendran2020}. %We consider a population with density-independent birth, death, and movement and also account for density-dependent birth and death processes. Specifically, individuals interact via binary reproductive facilitation and ternary competition. 
We consider a population with density-dependent birth and death rates due to the effect of facilitation and intraspecific competition respectively. Reproductive facilitation is typical even in species with asexual reproduction when individuals need the presence of conspecifics to reach the physiological condition to reproduce \citep{courchamp2008allee}. Some examples of species exhibiting asexual reproduction and reproductive facilitation are self-fertile snails and parthenogenetic female lizards \citep{Crews1986, Thomas1974}. Intraspecific competition, on the other hand, reduces individual fitness at very high population densities and is necessary to avoid unbounded population growth. Following standard notation in stochastic population dynamics \citep{doering2003,toral2014stochastic,constable2016,mendez2019demographic}, we can summarize the previous processes and interactions in the following set of demographic reactions
\begin{subequations}\label{eqs:nonspacerates}
\begin{eqnarray}
    && \emptycirc[0.9ex] \xrightarrow{b} \emptycirc[0.9ex] + \emptycirc[0.9ex]\label{eq:basebirth} \\[1ex]
    && \emptycirc[0.9ex] \xrightarrow{d} \text{\O}\label{eq:basedeath}\\[1ex]
    && \emptycirc[0.9ex] +  \emptycirc[0.9ex] \xrightarrow{\beta}  \emptycirc[0.9ex] + \emptycirc[0.9ex] + \emptycirc[0.9ex]\label{eq:coop}\\[1ex]
    && \emptycirc[0.9ex] + \emptycirc[0.9ex] + \emptycirc[0.9ex] \xrightarrow{\gamma}  \emptycirc[0.9ex] + \emptycirc[0.9ex] \label{eq:comp}
\end{eqnarray}
\end{subequations}
where (\ref{eq:basebirth}) and (\ref{eq:basedeath}) represent base-line reproductive success and death apart from intraspecific competition respectively; (\ref{eq:coop}) %represents a binary cooperative interaction 
describes the positive feedback in the reproduction rate (fitness gain) due to facilitation as an interaction between two individuals that results in the birth of a third organism at rate $\beta$. %in which two individuals interact at rate $\beta$ and produce a third individual
The last reaction, \eqref{eq:comp}, is a ternary competition process describing fitness losses due to, for example, crowding or resource depletion effects. This set of processes is one of the mathematically simplest ways of modeling a component Allee effect at the individual level \citep{mendez2019demographic}. However, one can think of many other density-dependent processes that might result in a component Allee effect, such as reduced death, sexual reproduction, or collective predation, among others \citep{drake2011allee,oro2020}. Any of these alternative processes can be incorporated into this modeling approach by simply modifying the set of reactions (\ref{eqs:nonspacerates}).

Combining the reactions in \eqref{eqs:nonspacerates} that contribute to the birth and the death of a new organism and using combinatorics arguments, we obtain the non-linear birth and death rates for a given population size $n$ \cite{nisbet1982,renshaw1993}, $\Omega(n\rightarrow n+1)$ and $\Omega(n\rightarrow n-1)$, respectively (see Supplementary Material S1):
\begin{eqnarray}\label{eq:nonspace-rates}
    \Omega(n\rightarrow n+1) &=& b\,n + \beta n(n-1) \\
    \Omega(n\rightarrow n-1) &=& d\,n + \gamma n(n-1)(n-2).
\end{eqnarray}
The difference between this birth and death rate results in a quadratic dependence of the per-capita reproduction rate on population size, similar to that reported by Allee in his experiments with isolated laboratory populations of the flour beetle \citep{allee1938social,allee1949principles}.

To introduce space explicitly in the model, we consider that individuals are located in the sites of a one-dimensional regular lattice with periodic boundary conditions, but it is straightforward to extrapolate the derivation to more realistic two-dimensional landscapes. We label each lattice node with an integer index $i\in[0,N]$ and denote the spatial coordinate with $x\in [0,L]$. The distance between two adjacent lattice nodes is $\delta x$ such that the spatial coordinate of the $i$-th node is $x_i = i \, \delta x$. Because organisms are now distributed in space, the state of the population is not fully determined by the total population size. Instead, we define a vector $\bm{\eta}$ that specifies the number of individuals in each spatial coordinate $\bm{\eta}=\{...n_{x-\delta x},\,n_{x},\,n_{x+\delta x}...\}$. 

In the lattice, organisms only interact with each other if they are within the interaction-specific range, $R_f$ and $R_c$ for facilitation and competition, respectively. As a result, the total birth and death rates depend on the density of organisms within these interaction ranges rather than on total population sizes as in Eq.\,\eqref{eq:nonspace-rates} (see Supplementary Material S1 for a full derivation using combinatorics). The total birth and death rate at each lattice coordinate $x$ thus become
\begin{eqnarray}
    \Omega\big(\bm{\eta}\to\{ n_x+1\}_{\bm{\eta}}\big) &=& b \,n_x + \dfrac{\beta \, n_x}{4 \, R_f}\left(N^{f}_x-1\right) \label{eq:totbirth-bl}\\
    \Omega\big(\bm{\eta}\to\{ n_x-1\}_{\bm{\eta}}\big) &=& d \,n_x + \dfrac{\gamma \, n_x}{24 R_c^2}\left(N^{c}_x-1\right)\left(N^{c}_x-2\right) \label{eq:totdeath-nl}
\end{eqnarray}
where $N^{f}_x$ and $N^{c}_x$ are the total number of organisms within a distance $R_f$ and $R_c$ from $x$ respectively. $\{n_x\pm 1\}_{\bm{\eta}}$ denotes a lattice configuration in which all nodes have the same number of individuals as in the configuration $\bm{\eta}$ except the node with spatial coordinate $x$, where the occupancy has changed in one unity. Notice that in Eqs.\,\eqref{eq:totbirth-bl}-\eqref{eq:totdeath-nl}, we are assuming that the non-local demographic rates do not depend on the distance between individuals as long as this distance is shorter than the interaction range. We are, therefore, modeling the interaction kernel with a top-hat function. The factors dividing the rates $\beta$ and $\gamma$ are normalizing factors of the top-hat kernel, and they make birth and death rates depend on population density rather than population size. Finally, we assume that individuals move on the lattice performing a nearest-neighbor random walk with rate $h$, which leads to a global transition rate due to movement
\begin{equation}\label{eq:jumps}
    \Omega\big(\bm{\eta}\to\{ n_x-1,n_{x'}+1\}_{\bm{\eta}}\big)=  h n_x.
\end{equation}
where \corr{$x'=x\pm\delta x$} and $\{n_x-1,n_{x'}+1\}_{\bm{\eta}}$ denotes a lattice configuration that only differs from $\bm{\eta}$ in the number of organisms at nodes $x$ and one of its nearest neighbors $x\pm\delta x$. This choice for the movement transition rates results in a diffusive movement with diffusion coefficient $D= h\, \delta x^2$.

Using the global transition rates in Eqs.\,\eqref{eq:totbirth-bl}-\eqref{eq:jumps} and also considering the movement dynamics, we can write the Master equation for the probability of finding a configuration $\bm{\eta}$ at time $t$ as
\begin{equation}\label{mastergen}
     \dfrac{\partial P(\bm{\eta},t)}{\partial t}= \sum_{\bm{\eta}'} \Omega(\bm{\eta}'\to \bm{\eta})P(\bm{\eta}',t) - \Omega(\bm{\eta}\to\bm{\eta}')P(\bm{\eta},t)
\end{equation}
\noindent where the sum in $\bm{\eta}'$ runs over all possible states.

\subsection{Derivation of the deterministic PDE approximation}
We use the Doi-Peliti formalism to derive a deterministic approximation of the spatial stochastic dynamics introduced in Section \ref{sec1:level1} \citep{doi1976,peliti1985,Tauber2007,hernandez2004clustering}. This deterministic approximation neglects demographic fluctuations and maps the set of discrete reactions to a PDE that describes the dynamics of a population density field $\rho(x,t)$ in continuous space and time. Hence, this approximation fails to describe noise-driven consequences of the Allee effect that might be ecologically relevant at low population sizes, such as extinctions caused by demographic noise \citep{mendez2019demographic}, or low diffusion scenarios. It, however, allows us to apply tools from spatially extended dynamical systems and obtain analytical insights of the underlying stochastic dynamics. More specifically, we can investigate in which conditions individuals form aggregates, and create a regular spatial pattern of population density \citep{Cross1993}. Following the steps detailed in the Supplementary Material section S1, the stochastic dynamics defined in Section \ref{sec1:level1} leads to the following partial differential equation for $\rho(x,t)$
\begin{equation}\label{eq:pderho}
    \dfrac{\partial \rho(x,t)}{\partial t} =\left[ r  + \beta \, \tilde{\rho}_f(x,t) - \gamma \, \left[\tilde{\rho}_c(x,t)\right] ^2\right]\rho(x,t) + D \,\nabla_x^2 \rho(x,t),
\end{equation}
where $r=b-d$ is the intrinsic growth rate and
\begin{eqnarray}
    \tilde{\rho}_\alpha(x,t)&=& \dfrac{1}{2 R_\alpha}\int_{x-R_\alpha}^{x+R_\alpha}  \rho\left(x^{\prime}, t\right) \mathrm{d} x^\prime
\end{eqnarray}
%\begin{equation}\label{G}
 %    G(\left|x-x^{\prime}\right|;R_\alpha)= \begin{cases}\frac{1}{2 R_{\alpha}} & \text { if } \left|x-x^{\prime}\right| \leq R_{\alpha} \\ 0 & \text { otherwise. }\end{cases}
%\end{equation}
with $\alpha=\{f,\,c\}$ for facilitation and competition, respectively. When the population density is uniform, the nonlocal model of Eq.\,(\ref{eq:pderho}) is mathematically equivalent to the cubic equation presented in the literature as the paradigmatic example of a population dynamics model with demographic Allee effect 
\citep{mendez2019demographic,kot2001ch2,oro2020ch6}. This cubic model has three stationary solutions. A trivial solution $\rho=0$ exists for any value of $r$ but is stable only for $r<0$. The other two possible stationary solutions, $\rho^{\s{U}}_{\pm}$ are the roots of the quadratic equation $r + \beta\, \rho - \gamma\, \rho^2=0$,
\begin{equation}\label{eq:solutions}
    \rho^{\s{U}}_{\pm} = \frac{\beta\pm\sqrt{\beta^2 + 4\gamma\, r}}{2\gamma}.
\end{equation}
\noindent
$\rho^{\s{U}}_{+}$ exists for $r>r^{\s{U}}_c=-\beta^2/4\gamma$ and is always stable, whereas $\rho^{\s{U}}_{-}$ exists for $r \in [r_c^{\s{U}}, 0]$ and is always unstable. Therefore, uniformly distributed populations are bistable for $r\in[r_c^{\s{U}},0]$, go extinct for $r < r_c^{\s{U}}$, and stabilize at $\rho^{\s{U}}_{+}$ for $r > 0$.

Finally, because $\rho(x,t)$ is a population density, we must integrate it over the system size to obtain the total population size,
\begin{equation}
    A (t)=\int_0 ^L \rho(x,t)dx.
\end{equation}
Therefore, at the stable equilibrium, we reach the abundance $\mathcal{A}=\lim_{t\to \infty} A(t)$.

\section{Results}
\subsection{IBM simulations and validation of the continuum approximation}\label{subsec:validation}
\begin{figure*}[ht]
\centering
\includegraphics[width={\linewidth}]{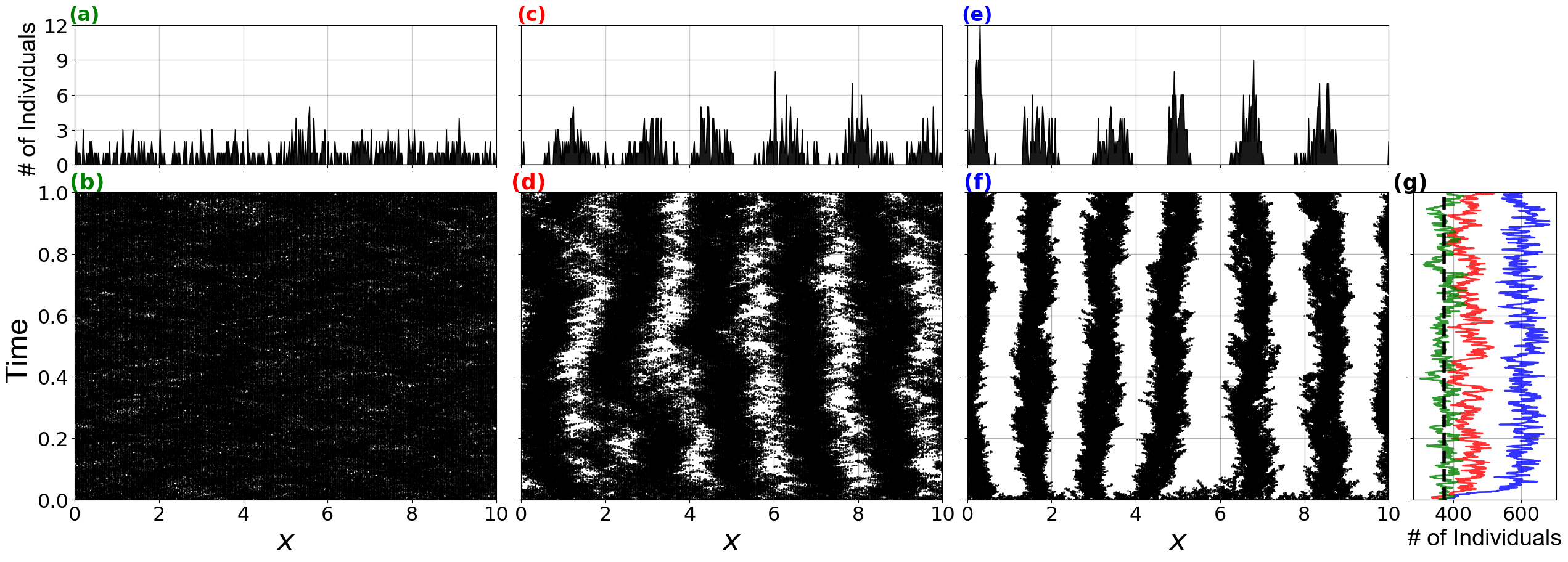}
\caption{\label{fig:2} IBM spatial patterns for high (a-b, $D=8$), intermediate (c-d, $D=1.2$), and low diffusion (e-f, $D=0.08$). Top panels (a, c, e) show the number of individuals at each lattice node at the end of a single simulation run. Bottom panels (b, d, f) show the temporal dynamics of the spatial distribution of organisms. The rightmost panel (g), shows the population size versus time at high (green), intermediate (red), and low (blue) diffusion together with the prediction from the non-spatial model (black-dashed line). Parameters (a-g): $b=30$, $d=40$, $\beta=4$, $\gamma=0.1$, $L=10$, $R_f=0.75$ and $R_c=1$, $\delta x=0.02$; uniform initial condition. See SM section S6.1 for details on the numerical methods.}
\end{figure*}

We first perform numerical simulations of the stochastic dynamics described by the Master equation \eqref{mastergen}, using the Gillespie algorithm \citep{Gillespie1977}. For high diffusion, i.e., high values of $h$, the population reaches a steady state with a uniform spatial distribution of organisms (Fig.\,\ref{fig:2}a,\,b). As diffusion decreases, however, individuals start to aggregate, and the population develops a spatial pattern characterized by isolated clumps of organisms interspersed with unpopulated regions (Fig.\,\ref{fig:2}c-f). Moreover, the total population size increases in the stationary state due to aggregation (Fig.\,\ref{fig:2}g), indicating that grouping improves the environmental conditions and increases the system carrying capacity. The same type of spatial structure and population dynamics are observed in two dimensions (Fig.\,S1). 

\begin{figure}[H]\centering
\includegraphics[width={\linewidth}]{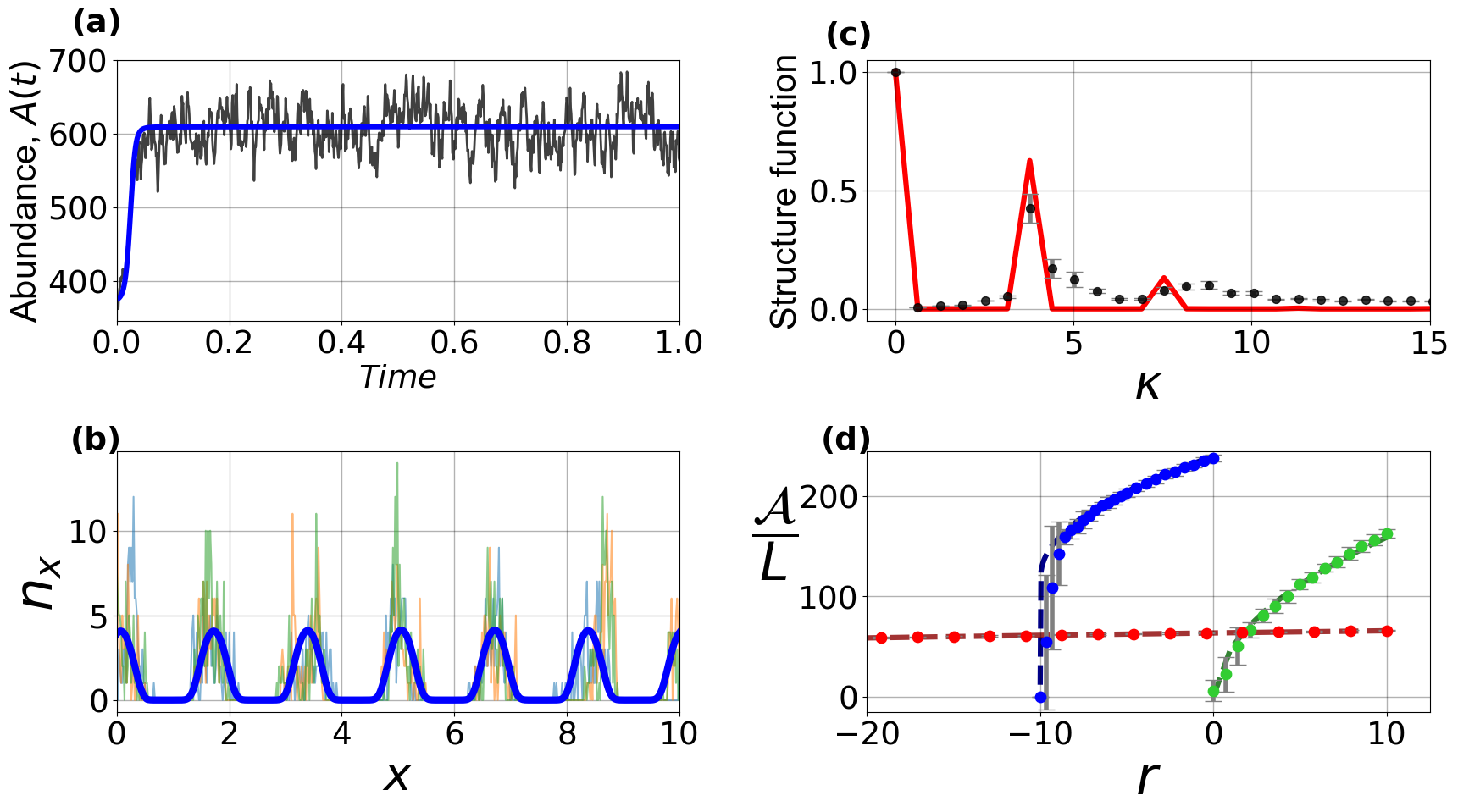}
\caption{\label{fig:3} Comparison between the IBM and its continuum approximation. a) Population size versus time for a single IBM realization (black line) and the deterministic approximation (blue). b) Fourier analysis of the IBM (black dots) and PDE spatial patterns (red line). For the IBM, we averaged the structure functions of $100$ spatial patterns evenly sampled in the interval $t\in[0.1,1]$. Error bars represent the variance in this dataset. c) IBM long-term spatial patterns at different times (thin blue, orange, and green lines) and their corresponding PDE approximation (blue thick curve). Parameters in (a)-(c) are the same used in Fig.\,\ref{fig:2}e,\,f. d) Long-term population size versus $r$ for populations exhibiting strong Allee effect (blue; $\beta=10^{-1}$, $\gamma=10^{-3}$, $D=10^{-3}$, $R_f=0.5$), no Allee effect (green; $\beta=0$, $\gamma=10^{-3}$, $D=10^{-3}$, $R_f=0.5$), and strong facilitation preventing extinctions (red; $\beta=4$, $\gamma=0.1$, $D=0.08$, and $R_f=0.75$). Solid lines correspond to PDE numerical results, and dots are averages of $50$ IBM simulations with error bars indicating the variance (see SM section S6 for details of the numerical simulations)}
\end{figure}

Next, we test the accuracy of the PDE approximation by comparing these simulation outcomes with numerical solutions of Eq.\,\eqref{eq:pderho}. This comparison returns an excellent quantitative agreement between the stochastic individual-level dynamics and the deterministic equation regarding the total population size and periodicity of the spatial patterns (Figs.\,\ref{fig:3}a,\,b). For the latter, we obtained the typical distance between aggregates using the pattern structure function, which can be calculated as the modulus of the spatial Fourier transform of the population density \cite{Martinez-Garcia2015,hernandez2004clustering}. We only find two points of disagreement between the IBM and its corresponding PDE approximation, both due to neglecting fluctuations in the latter. First, the centers of the aggregates change with time in the IBM simulations, whereas they are stationary in the PDE (Fig.\,\ref{fig:3}c). Second, for $r$ values close to $r_c$, fluctuations in the IBM simulations can drive the population size below the Allee threshold and cause extinctions, which makes the average population density for these values of $r$ differ between IBM and PDE simulations \citep{dennis2002} (Fig.\,\ref{fig:3}d, blue and green curves). %Thus, fluctuations become an important driver of population dynamics in this parameter regime, and the mean-field results diverge from the stochastic ones. 
Besides these two features induced by fluctuations, the population-level model captures the main spatial and demographic phenomenology observed in the IBM simulations. This excellent agreement between both modeling approaches allows us to use the PDE approximation to investigate in what conditions aggregates form and their population-level consequences.

\subsection{Group formation}\label{subsec:groupform}
\begin{figure}[ht]\centering
\includegraphics[width={0.8\linewidth}]{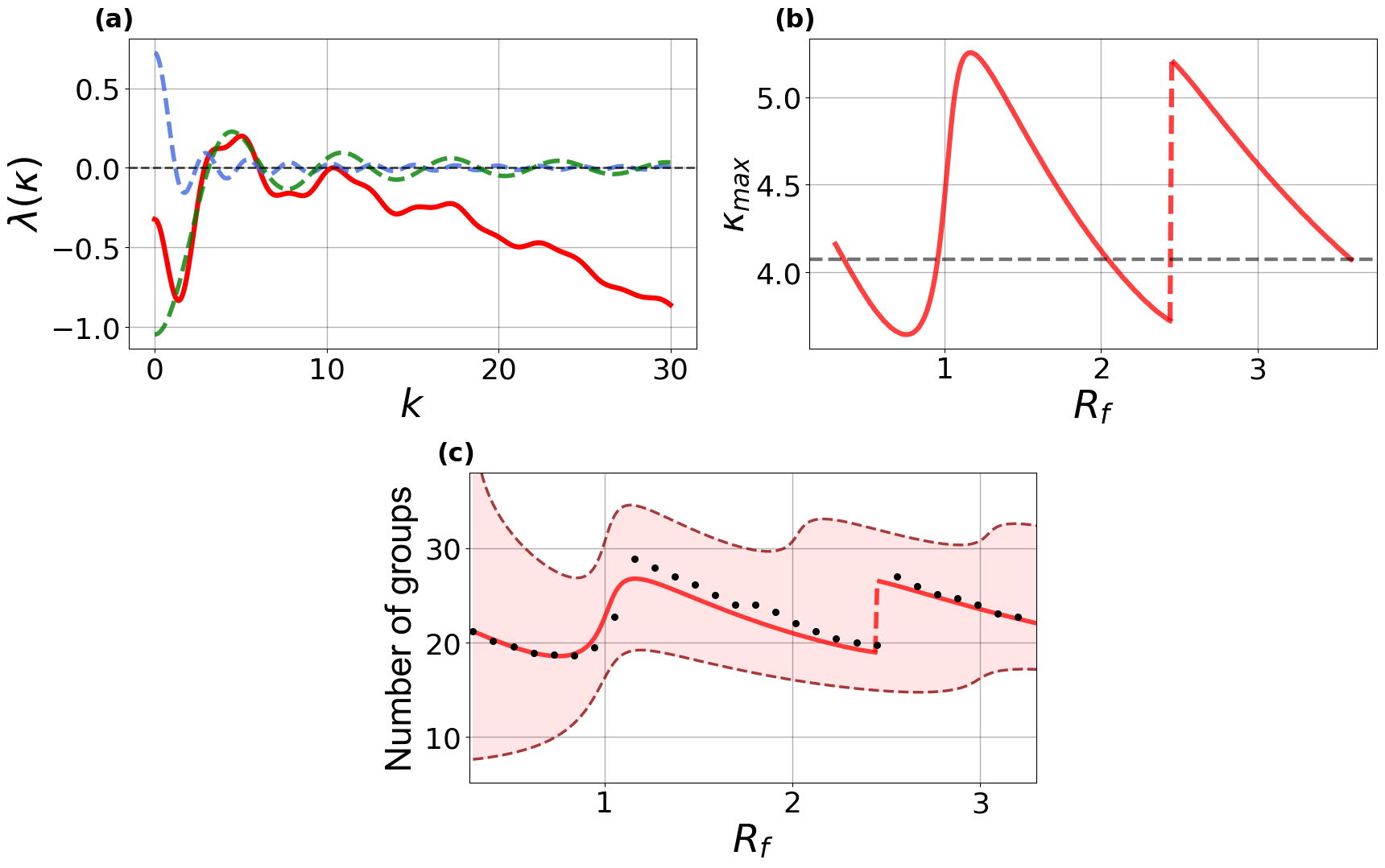}
\caption{\label{fig:4} a) Perturbation growth rate as a function of the wavenumber $k$ (red). The dashed lines represent the contributions of the facilitation (blue) and competition (green) terms to $\lambda(k)$. b) Fastest growing wavenumber, $k_{\s{max}}$ versus $R_f$. The grey dashed line is $k_{\s{max}}$ in the limit $\beta \to 0$. c) Number of peaks predicted by the linear stability analysis (red line and shaded region) and observed numerically in the deterministic equation (black dots). The shaded red region accounts for uncertainty in the number of peaks due to the existence of several wavenumbers with positive growth rates. For panels b and c, the dashed red lines represent discontinuities in $k_{\s{max}}$ and number of peaks. Parameter values: $r=-2$, $D=0.001$, $\beta=1$, $\gamma=1$ and, $L=32$, $R_c=1$; $R_f=2.6$ (in panel a).}
\end{figure}

To investigate in which conditions organisms aggregate and determine the key ingredients for pattern, we perform a linear stability analysis of Eq.\,\eqref{eq:pderho} \citep{Cross1993}. This technique consists in adding a small spatial perturbation to a stable uniform solution of the equation and calculating the perturbation growth rate. If the perturbation growth rate is negative, the uniform solution is stable and patterns do not form. Conversely, the perturbation leads to spatially periodic solutions or patterns if its growth rate is positive \citep{Cross1993}. We consider a solution of the form $\rho(x,t) = \rho^{\s{U}}_+ + \epsilon \psi(x,t)$ where $\rho^{\s{U}}_+$ is the uniform non-zero solution of Eq.\,\eqref{eq:pderho}, and $\psi(x,t)$ is an arbitrary perturbation modulated by an amplitude parameter $\epsilon \ll1$. We insert this solution into Eq.\,(\ref{eq:pderho}) and obtain a partial differential equation for the dynamics of the perturbation $\psi(x,t)$. By linearizing and Fourier transforming this partial differential equation, we obtain the perturbation growth rate as a function of its wavenumber $k$ (see Supplementary Material section S2 for details of the calculation; Fig.\,\ref{fig:4}a). This perturbation growth rate is
\begin{equation}\label{eq:pertgr}
    \lambda(k) = \rho^{\s{U}}_+ \Big[ \beta \dfrac{\sin(kR_f)}{kR_f} -2 \gamma \rho^{\s{U}}_+ \dfrac{\sin(kR_c)}{kR_c}  \Big ] - D k^2.
\end{equation}

\noindent If $\lambda(k)$ is positive for a given wavenumber $k$, a perturbation with that wavenumber will grow and create a regular pattern of population density. The wavenumber maximizing $\lambda(k)$ in Eq.\,\eqref{eq:pertgr}, $k_{\s{max}}$, defines the dominant periodicity of the spatial pattern at short times and is related to the periodicity of the long-term spatial pattern of population density (Fig.\,\ref{fig:4}b, where we find an excellent agreement between $k_{\s{max}}$ for $R_f=0.75$ in Fig.\,\ref{fig:4}b and the peak of the structure functions in Fig.\,\ref{fig:3}b). Hence, we can estimate the number of groups $m$ that form in a system of size $L$ as $m\approx L\,k_{\s{max}}/2\pi$ (Fig.\,\ref{fig:4}c). Moreover, we can better understand how the different processes and interactions included in the microscopic model contribute to pattern formation by analyzing term by term all the different contributions to the perturbation growth rate.

\begin{figure}[ht]\centering
\includegraphics[width={0.5\linewidth}]{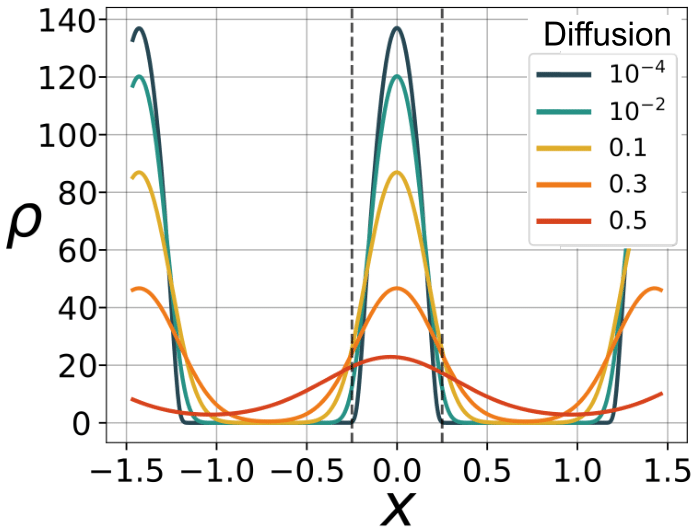}
\caption{\label{fig:diff} Effect of diffusion on the long-term patterns of population density. The vertical black dashed lines limit the extent of the facilitation range, $R_f$, for organisms located at $x=0$. Parameter: $t=2\times 10^{4}$, with $dt=0.05$, $dx=0.008$, $L=32$ and parameters: $r=-2$, $R_f=0.5$ and $R_c=1$. See Supplementary Material sections S3 and S6 for details on the numerical methods.}
\end{figure}

First, the linear stability analysis shows that diffusion contributes with a negative term to Eq.\,\eqref{eq:pertgr} and tends to homogenize population density and eliminate patterns (Fig.\,\ref{fig:diff}). Second, long-range competition and facilitation enter the perturbation growth rate via the Fourier transform of their corresponding interaction kernel, which, in the case of the top-hat kernel chosen in the model, are damped oscillatory functions with interaction-specific frequency, amplitude, and sign (dashed lines in Fig.\,\ref{fig:4}a). The interaction range determines the frequency of each oscillatory function, while the intensity of the intraspecific interaction determines the amplitude of the oscillations. The sign preceding each oscillatory function indicates how competition or facilitation impacts population growth, with the negative sign corresponding to competition and the positive one to facilitation.

To better understand the role of long-range competition and facilitation in the formation of aggregates, we next consider the limit cases in which diffusion is low and each of these interactions vanishes or acts on a local scale. In the local competition limit, $R_c \to 0$, the perturbation growth rate is always negative because $\rho_+^{\s{U}} < \beta/(2\,\gamma)$ when populations are uniformly distributed, i.e. $\lambda(k_{\s{max}})<0$, [see Eq.\,\eqref{eq:solutions}]. Therefore, patterns do not form. However, if facilitation is local, $R_f\to 0$, or vanishes, $\beta = 0$, the perturbation growth rate can still be positive for certain wavenumbers, and patterns can potentially form. Because in this low diffusion limit $R_f$ defines the aggregate width (gray-dashed lines in Fig.\,\ref{fig:diff}), varying the facilitation range makes the fastest-growing wavenumber, and, therefore, the number of groups oscillate around the value obtained when long-range interactions are purely competitive. Therefore, long-range competition is the key non-locality for pattern formation, and it sets the periodicity of the long-term spatial pattern of population density. Conversely, facilitation plays a secondary role in pattern formation, rearranging the pattern periodicity around the value set by the competition range \cite{rietkerk2008regular}. Previous studies have already identified long-range competition as a cause of spatial patterns through the establishment of the so-called exclusion regions, i.e., regions between clusters of organisms in which individuals would compete with individuals from two neighbor groups \citep{hernandez2004clustering,martinez2013vegetation,martinez2014minimal,martinezgarcia2022}. Moreover, the spatial patterns of population density exhibit aggregates shorter than the range of both non-local interactions, which makes the intensity of competition and facilitation inside an aggregate approximately constant (see vertical black dashed lines in Fig.\,\ref{fig:diff}).

%Finally, we computed the stationary-state population size as a function of the diffusion coefficient to evaluate the range of diffusion intensity at which the first two assumptions underlying the group-level approximation in Eq.\,\eqref{met} remain valid (Fig.\,\ref{fig:8}a). Consistently with the simulations of the stochastic individual-based dynamics (Fig.\,\ref{fig:2}), we observe that the total population abundance decreases as diffusion increases. In the low-diffusion regime, the population abundance agrees with the predictions of the meta-population approximation, \corr{and the spatial scales of peak size and exclusion region sizes approach $R_f$ and $R_c$, in accordance with the group-level model assumptions}. However, as diffusion increases, diffusion takes control of the spatial dynamics, and the assumptions underlying the metapopulation approximation stop being valid. As a result, the population density decreases until diffusion reaches a critical value (\corr{vertical} black dashed line in Fig.\,\ref{fig:8}a), at which patterns do not form, and the population abundance is equal to that predicted by models assuming uniformly distributed individuals. We also observe this decrease in population density in the spatial patterns of population density, which tend to become uniform as diffusion increases (Fig.\,\ref{fig:8}b).

\subsection{The effect of spatial patterns on the demographic Allee effect}\label{pop_dist}
In the previous section, we investigated the conditions in which organisms form non-uniform patterns of population density and quantified the features of the emergent aggregates. Next, we study how aggregation impacts the demographic Allee effect compared to a uniformly distributed population. More specifically, we focus on how group formation affects the main features of a strong demographic Allee effect: the stationary population density, the Allee threshold, and the value of the net growth rate at which extinction is the only stationary state, $r_c$. Because we are interested in the strong-Allee-effect regime, we limit the analysis to negative density-independent net population growth rates, $r<0$. In this parameter regime, for uniform population densities, $r_c^{\s{U}}=-\beta^2/(4\,\gamma)$ and both $\rho_+^{\s{U}}$ and $\rho_-^{\s{U}}$ exist and are positive for $r\in [r_c^{\s{U}}, 0]$. This range of values of $r$ defines the parameter regime in which a uniformly distributed population exhibits a strong demographic Allee effect, with Allee threshold equal to $\rho_{-}^{\s{U}}$ and stationary population density $\rho_+^{\s{U}}$.

As we observed in the IBM simulations (Fig.\,\ref{fig:2}), aggregation increases the stationary population density and allows for population survival at $r<r_c^{\s{U}}$ (Fig.\,\ref{fig:6}a,\,b). Moreover, the Allee threshold becomes space-dependent and is determined by the local density of individuals within the competition and facilitation ranges \cite{boukal2002}. These local densities, in turn, depend on the number and spatial arrangement of groups. As a result of these changes in $r_c$ and the Allee threshold, aggregation allows populations subjected to a component Allee to persist in harsher environments and at higher numbers than those uniformly distributed \cite{kao2023,Surendran2020}. Moreover, because spatially structured populations have lower Allee thresholds, they are less susceptible to extinctions caused by environmental perturbations and can recover after extinction following smaller fluctuations than uniformly distributed populations. We obtained these results using the deterministic approximation in Eq.\,\eqref{eq:pderho}, which allows us to compute both stable and unstable solutions of the model (see Supplementary Material section S3 for a detailed description of how to obtain the bifurcation diagram in Fig.\,\ref{fig:6}). We further tested these predictions with direct numerical simulations of the individual-level stochastic dynamics and obtained an excellent agreement for most values of $r$, as already shown in Section \ref{subsec:validation} (Fig.\,\ref{fig:3}c).

\subsubsection{Analytical results for the demographic Allee effect: a metapopulation approximation}

To develop a more mechanistic understanding of how spatial patterns impact the properties of the demographic Allee effect, we further approximate the deterministic equation (\ref{eq:pderho}) for population density by a network, metapopulation-like description in which each node or population represents a group of individuals and each link represents the existence of inter-group facilitation. We build this approximation based on three features of the spatial patterns of population density. First, all individuals within a group must interact with one another via competition and facilitation. Mathematically, this means that competition and facilitation ranges must be greater than clusters of organisms. Second, individuals from different groups must not compete with each other. In terms of the model equation, this condition implies that the competition range must be shorter than the distance between pattern aggregates. Finally, if two groups interact with each other via facilitation, this positive interaction must reach all the individuals in both groups. Therefore, the facilitation range must be large enough to encompass all the individuals of a neighbor group. The first two assumptions are only met when diffusion is low and the non-local ecological interactions determine the population spatial structure. The last assumption is correct if the first two are met, except for specific values of $R_f$ for which the facilitation range partially reaches neighboring clusters. When these three conditions are fulfilled, the nonlocal densities in Eq.\,\eqref{eq:pderho} are constant inside each aggregate, and the integral defining them reduces to the number of individuals within each group $\mathcal{N}_I$. Using this simplification for the convolution terms in Eq.\,\eqref{eq:pderho} and integrating over the group size on both sides of the equation, we obtain an equation for the dynamics of aggregate size (see Supplementary Material section S4): 
\begin{equation}\label{met_e}
     \dfrac{\partial \mathcal{N}_I(t)}{\partial t} = \Bigg[ r + \dfrac{\beta}{2 R_f}\left(\mathcal{N}_I(t)+\sum_{\langle I,J\rangle} \mathcal{N}_J(t)\right) - \gamma \,  \dfrac{\mathcal{N}_I^2(t)}{4 R_c ^2} \Bigg]\mathcal{N}_I(t)
\end{equation}
where "$\mathcal{N}_I$" is the number of individuals in the focal aggregate "$I$" and "$J$" is an index that runs over the neighbors of $I$ that are within the facilitation range. Here, the spatial properties of the pattern are stationary, and the state of the population is fully determined by the size of each aggregate.% 

Since there are no spatial dependencies in the parameters, and considering periodic boundary conditions, we can assume that all aggregates have the same number of neighbors and the same population size $\mathcal{N}$ in the stationary state. In these conditions, we can introduce a parameter that gives the number of groups within the facilitation range, $\xi$, which depends on both the facilitation range and the pattern wavelength. Using this parameter, we can write an ordinary differential equation to describe the dynamics of any aggregate size give a fixed spatial distribution of groups:
\begin{equation}\label{met}
     \dfrac{\partial \mathcal{N}(t)}{\partial t} = \Bigg[ r + \beta  (\xi+1)\, \dfrac{\mathcal{N}(t)}{2 R_f} - \gamma \,  \dfrac{\mathcal{N}^2(t)}{4 R_c ^2} \Bigg]\mathcal{N}(t).
\end{equation}

 This equation encodes all the information about the underlying network of inter-group interactions in the parameter $\xi$. Solving Eq.\,\eqref{met} we can obtain the possible stationary group sizes: $\mathcal{N}_0=0$ (extinction) and
\begin{equation}\label{N}
    \mathcal{N}_{\pm} =   \dfrac{(\xi + 1 )\dfrac{\beta}{2R_f} \pm \sqrt{\Bigg ((\xi + 1 ) \, \dfrac{\beta}{2R_f} \Bigg )^2 +\dfrac{r \gamma}{R_c^2}}}{\gamma/ 2R_c ^2}.
\end{equation}

\begin{figure}[ht]\centering
\includegraphics[width={\linewidth}]{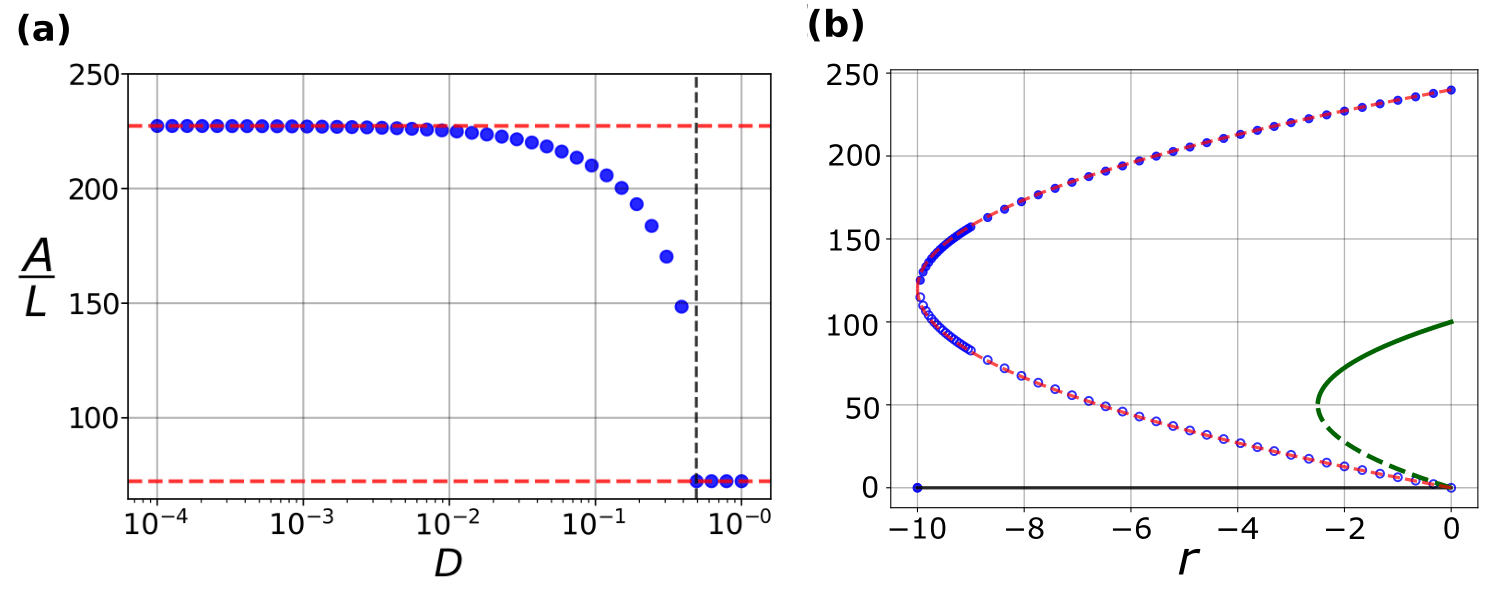}
\caption{\label{fig:6}(a) Population abundance versus diffusion obtained from the PDE approximation (blue dots). The red dashed lines show the metapopulation approximation, $\mathcal{N}_+$ in Eq.\,\eqref{N} (top), and the uniform solution,  $\rho_+$ in Eq.\,\eqref{eq:solutions} (bottom). The vertical gray dashed line indicates the critical diffusion at which patterns disappear. (b) Population abundance versus net population growth rate, $r$, obtained from pattern-forming solutions of the PDE approximation, Eq.\,\eqref{eq:pderho} (blue circles); the non-spatial cubic model (green curves); and the group-level approximation, Eq.\,\eqref{N} (dashed red line). The filled circles and the green solid line represent a stable equilibrium, whereas the empty symbols and green dashed lines represent unstable equilibrium states. The deterministic simulations run until $t=7.5$, with $dt=0.005$ and $dx=0.008$, $\beta=10^{-1}$, $\gamma=10^{-3}$, $R_f=0.5$, $R_c=1$ and $\delta x=0.02$. All simulations are done with $L=32$ and $D=10^{-3}$. See Supplementary Material sections S3 and S6 for details on the numerical methods.}
\end{figure}

The predictions of this meta-population approximation for $r_c$ and the steady-state population size are in excellent agreement with those of the density equation and the outcome of the stochastic simulations. In addition, mapping the spatially explicit dynamics to a set of coupled ordinary differential equations allows us to obtain analytical expressions for these two features of the demographic Allee effect in the presence of spatial patterns of population density. The steady-state population size is $A=m\,\mathcal{N}_+$, where $\mathcal{N}_+$ is given by Eq.\,(\ref{N}) and $m$ is the number of groups that we can estimate from the pattern wavelength predicted by the wavenumber that maximizes the perturbation growth rate in Eq.\,(\ref{eq:pertgr}), $k_{\s{max}}$ (red lines in Fig.\,\ref{fig:6}). Imposing $\mathcal{N}_+ = \mathcal{N}_-$ in Eq.\,\eqref{N}, we can calculate the critical value of the net growth rate that can sustain a non-zero population size,
\begin{equation}\label{eq:rc}
    {r}_c= - \gamma^{-1} \bigg[ \frac{\beta}{2}\frac{R_c}{R_f}\,(\xi + 1) \bigg]^2.
\end{equation}

As expected, $r_c$ decreases with increasing facilitation and decreasing competition strength $\beta$ and $\gamma$, respectively. Moreover, because the pattern wavelength is an oscillatory function of the interaction ranges, $r_c$ can either increase or decrease when increasing $R_f$ and $R_c$. This complex dependence makes that, for certain net growth rates $r$, a population will only survive if groups facilitate each other, which shows that increasing facilitation range can sustain populations in certain environmental conditions. Notice, however, that when the facilitation range increases and groups rely on one another for survival, the whole population becomes less resistant to local perturbations that might cause global extinctions due to the high connectivity between groups.

\subsubsection{Implications for the group Allee effect}
\begin{figure}[ht]\centering
\includegraphics[width={0.6\linewidth}]{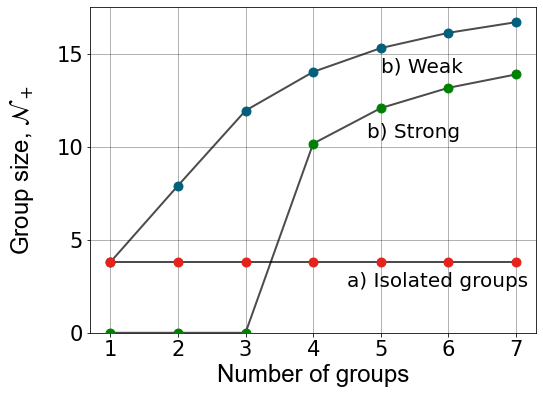}
\caption{\label{fig:7} Demographic Allee effect in a population composed of several groups. Here, we set the number of groups in the system and compute the size of a single group in the stationary state, $\mathcal{N}_+$. The red dots correspond to a situation in which groups are isolated, $r=-2$ and $R_f =0.5$; blue and green dots correspond to cases with inter-group facilitation with $R_f=2$ and $r=-2$ (blue) and $r=-100$ (green). For all cases, $R_c=1$, $\beta=10^{-1}$, and $\gamma=10^{-3}$}
\end{figure}

Organism grouping sets new ways in which the individual-level component Allee effect manifests at the population level and determines the Allee threshold. We analyze these possible outcomes for different numbers of groups and facilitation ranges using the meta-population approximation in Eq.\,\eqref{met_e} that gives the dynamics of each group independently. Mimicking the one-dimensional landscape we used in all previous analyses, we consider that groups are arranged in a line. However, we do not consider periodic boundary conditions to prevent the number of groups from being effectively infinite. If the facilitation range is short so individuals in different groups do not interact with one another, the fitness of the individuals within each aggregate only depends on group size (Fig.\,\ref{fig:7}a), and groups are independent units. Consequently, the formation or extinction of a group does not have any effect on the others, and the minimum population size that ensures population survival is equal to the Allee threshold of one single group, $\mathcal{N}_{-}$ from Eq.\,(\ref{N}) with $\xi=0$. In this limit, we recover the classic Allee effect cubic model for each group, which has been termed in the literature as the group Allee effect, i.e., a facilitation-related fitness component increases only with group size rather than total population size/density \citep{lerch2018demographic, angulo2013social, courchamp2008allee}.

When the facilitation range increases and is such that groups interact with one another, the fitness of the individuals can increase significantly due to the presence of neighbor groups. As a consequence, group size increases in the presence of more groups (Fig.\,\ref{fig:7}b and \ref{fig:7}c), and the Allee threshold is $\mathcal{N}_{-}$ from Eq.\,(\ref{N}) with $\xi>0$. For very harsh environmental conditions (low $r$), the population only survives if groups facilitate one another (Fig.\,\ref{fig:7}c).

\section{\label{sec3} Discussion}

We introduced a theoretical framework to investigate the demographic consequences of a component Allee effect across various levels of spatial organization as illustrated in Fig.\,\ref{fig:0}. This framework incorporates a component Allee effect arising from reproductive facilitation, which increases the reproduction rates of a focal individual with the population density within its neighborhood and growth limitation caused by intraspecific competition. Extending the analysis to other types of individual-level interactions leading to component Allee effects, such as social behaviors, mate limitation, environmental conditioning, or density-dependent movement should be straightforward \citep{courchamp2008allee,Fronhofer2015,oro2020}. We focused on quantifying the impact of the spatial distribution of organisms on specific features of the demographic Allee effect, such as the Allee threshold, the long-term total population size, and the lowest value of the density-independent growth rate for which the population survives. We measured these quantities in both uniformly and non-uniformly distributed populations. 

Similarly to Surendran \textit{et al.,} 2020 \cite{Surendran2020}, the framework we presented in this study differs from non-spatial and spatially implicit metapopulation models by explicitly considering the range of interaction for both reproductive facilitation and crowding effects. This level of detail is partially captured by metapopulation models, which assume that individuals only interact with others within the same group \cite{lerch2018demographic}. Metapopulation frameworks, however, assume a fixed population structure in groups, whereas we describe group formation dynamically as a result of individual-level processes. This explicit description of group formation allows us to identify the individual-level processes that control for each of the population-level features of the demographic Allee effect and subsequently manipulate them to understand how different spatial structures impact the demographic Allee effect. 

In addition, limiting the mechanisms responsible for the component Allee effect to a finite neighborhood around each focal individual makes the population dynamics and the features of the emergent demographic Allee effect depend on local, instead of total, population densities. For example, the Allee threshold becomes a local feature of the population that depends on the density of individuals within a given region of the landscape. This feature of the Allee threshold might enable the survival of local populations in situations where the global density is very low, which is especially relevant when spatial fluctuations in population density are high, such as in the presence of clumps of organisms or groups \citep{kanarek2013allee}. This strong dependence of the Allee threshold on the spatial population structure might help to explain field studies reporting population survival at low global population densities \citep{woodroffe2011demography, lundquist2011estimating, rijnsdorp2001feeding}.

The framework we establish here also provides the appropriate theoretical framework to formalize the group Allee effect and integrate it within a unifying modeling approach \citep{angulo2013social,angulo2018allee}. When organisms aggregate, one can consider the groups as the fundamental units of the population. If competition acts on a longer range than facilitation, these groups are independent units that do not interact with one another. In consequence, the component Allee effect impacts the demographics of a single group, resulting in a demographic group Allee effect that only determines the population dynamics within that group. This same argument can be extended to cases in which facilitation acts on a longer range than competition. In this limit, groups interact with one another, which can result in a group-level Allee effect when the fitness of a group increases in the presence of neighbors. This group-level component Allee effect scales to the population level by creating an emergent demographic Allee effect acting on groups that can even result in the existence of a minimum number of groups to ensure population survival.

Beyond group-level processes, non-uniform population densities favor population survival when the density within a region of the landscape is locally above the Allee threshold. Moreover, for the specific processes considered here, groups form to minimize competition. Therefore, aggregation minimizes competition and results in larger global population sizes that are less prone to extinction due to demographic fluctuations \citep{dennis2002}. This mechanism could explain why, across different taxa, several species form larger groups in response to worsening environmental conditions \cite{kao2023}. Aggregation also lowers the Allee threshold significantly, which favors the persistence of local populations at lower population densities. This local decrease in the Allee threshold is different from the effective decrease in the global Allee threshold discussed above, which is related to the locality of the Allee threshold rather than to its value. This difference between local and global Allee thresholds emphasizes the importance of considering the several scales involved in component and demographic Allee effects \citep{boukal2002}.

The bottom-up approach to study Allee effects we present here also explains mechanistically why aggregation enables population survival in harsher environments than uniform population densities \citep{Surendran2020,kanarek2013allee,kao2023}, and it allowed us to derive a mathematical expression relating this difference in the critical net population growth rate to the intensity and characteristic spatial scale of each of the individual-level processes governing population dynamics \citep{boukal2002}. Finally, we show how groups disappear at higher movement rates and induce an abrupt loss in population size. The detrimental impact of high dispersal on species exhibiting Allee effects is well-known in the literature, particularly in the context of species invasion \citep{lewis1993allee, kanarek2013allee}. We extend those results and show that stationary population sizes are also very sensitive to dispersal, which suggests that component Allee effects might impose strong selection pressures on dispersal \citep{Heinsalu2013,Pigolotti2014,Shaw2015}.

Aiming to provide the simplest framework to study Allee effects across levels of spatial organization, %and a unifying theoretical approach to understanding how Allee effects operate for different population structures. To keep it as simple as possible, 
we made some simplifying modeling assumptions. The choice of the component Allee effect, as we discussed before, can be easily changed by modifying the set of individual-level demographic reactions. Other assumptions, such as the choice of the interaction kernels, would not change our results provided that they lead to spatial pattern formation \citep{pigolotti2007,martinez2013vegetation,colombo2022,colombo2023pulsed}. These results are also insensitive to changes in the mechanism responsible for spatial pattern formation if spatial patterns emerge in the form of clumps of population density. We considered non-local interactions because it is the most straightforward way to create aggregation patterns \citep{martinez2014minimal}. An interesting direction for future research, however, would be to consider alternative pattern-forming interactions, such as density-dependent movement or resource-consumer interactions, leading to a larger variety of spatial patterns in population density, such as labyrinths and gaps \citep{liu2013,rietkerk2008regular,rao2016,Martinez-Garcia2015,Martinez-Garcia2022}. Finally, this modeling framework is also easily extendable to include interactions between several species \citep{maciel2021,simoy2023}, thus providing a theoretical tool to investigate community-level consequences of different component Allee effects.

In summary, this work showed three ways in which aggregation can influence population dynamics in the presence of Allee effects. First, aggregation enhances population density locally and thus allows the population to persist in harsh environments where uniformly distributed individuals would go extinct. Second, aggregation results in localized sub-populations that follow independent dynamics from one another and might eliminate the population-level Allee effect. Finally, aggregation decreases competition by limiting its effect to individuals within the same group. Consequently, aggregation reduces the Allee threshold, increases the total population size, and aids population survival in harsher environmental conditions. More generally, our work emphasizes the importance of bottom-up models developed from a rigorous description of the individual-level interactions and processes to develop a mechanistic understanding of observed patterns and trends in population dynamics.

\section*{Author Contributions}
DJ and RMG conceived the project and designed the model; DJ led all the steps of the model implementation, analysis and visualization with input from RMG; DJ and RMG wrote the manuscript. All authors gave final approval for paper publication.

\section*{Acknowledgments}
The authors thank all the reviewers for their time and attention spent on the manuscript.

\section*{Data Accessibility}
All the scripts needed to replicate the results presented here are available at \cite{zenodo}

\section*{Funding Statement}
This work was partially funded by the Center of Advanced Systems Understanding (CASUS), which is financed by Germany’s Federal Ministry of Education and Research (BMBF) and by the Saxon Ministry for Science, Culture and Tourism (SMWK) with tax funds on the basis of the budget approved by the Saxon State Parliament. Additional funding was provided by FAPESP through a Master Fellowship no.\,2020/15643-8 (DCPJ), a BIOTA Young Investigator Research Grant no.\,2019/05523-8 (RMG and DCPJ), and ICTP-SAIFR grant no.\,2016/01343-7; the Abdus Salam ICTP through the Associate's Programme, and the Simons Foundation through grant no.\, 284558FY19 (RMG).

\bibliography{bibliography}% Produces the bibliography via BibTeX.

\begin{thebibliography}{10}

\bibitem{stephens1999allee}
Stephens PA, Sutherland WJ, Freckleton RP.
\newblock What is the {A}llee effect?
\newblock Oikos. 1999:185-90.

\bibitem{levitan2005}
Levitan DR.
\newblock The {A}llee effect in the sea.
\newblock Marine {C}onservation {B}iology: the science of maintaining the sea's biodiversity. 2005:47-57.

\bibitem{courchamp2008allee}
Courchamp F, Berec L, Gascoigne J.
\newblock {A}llee effects in ecology and conservation.
\newblock OUP Oxford; 2008.

\bibitem{lande1987extinction}
Lande R.
\newblock Extinction thresholds in demographic models of territorial populations.
\newblock The American Naturalist. 1987;130(4):624-35.

\bibitem{sun2016mathematical}
Sun GQ.
\newblock Mathematical modeling of population dynamics with {A}llee effect.
\newblock Nonlinear Dynamics. 2016;85(1):1-12.

\bibitem{oro2020ch6}
Oro D.
\newblock Extinction, Nonlinear Dynamics, and Sociality.
\newblock In: Perturbation, behavioural feedbacks, and population dynamics in social animals: when to leave and where to go. Oxford University Press, USA; 2020. p. 114-27.
\newblock doi:10.1093/oso/9780198849834.003.0007.

\bibitem{berec2007}
Berec L, Angulo E, Courchamp F.
\newblock Multiple Allee effects and population management.
\newblock Trends in Ecology \& Evolution. 2007;22(4):185-91.
\newblock doi:10.1016/j.tree.2006.12.002.

\bibitem{orr2009fitness}
Orr HA.
\newblock Fitness and its role in evolutionary genetics.
\newblock Nature Reviews Genetics. 2009;10(8):531-9.
\newblock doi:10.1038/nrg2603.

\bibitem{drake2011allee}
Drake J, Kramer A.
\newblock {A}llee effects.
\newblock Nature Education Knowledge. 2011;3(10):2.

\bibitem{allee1932studies}
Allee W, Bowen ES.
\newblock Studies in animal aggregations: mass protection against colloidal silver among goldfishes.
\newblock Journal of Experimental Zoology. 1932;61(2):185-207.
\newblock doi:10.1002/jez.1400610202.

\bibitem{stephens2002model}
Stephens PA, Frey-roos F, Arnold W, Sutherland WJ.
\newblock Model complexity and population predictions. The alpine marmot as a case study.
\newblock Journal of Animal Ecology. 2002;71(2):343-61.
\newblock doi:10.1046/j.1365-2656.2002.00605.x.

\bibitem{allee1949group}
Allee W, Rosenthal G.
\newblock Group survival value for \textit{Philodina rosola}, a rotifer.
\newblock Ecology. 1949;30(3):395-7.

\bibitem{ghazoul2005buzziness}
Ghazoul J.
\newblock Buzziness as usual? {Q}uestioning the global pollination crisis.
\newblock Trends in ecology \& evolution. 2005;20(7):367-73.

\bibitem{dechmann2010group}
Dechmann DK, Kranstauber B, Gibbs D, Wikelski M.
\newblock Group hunting—a reason for sociality in molossid bats?
\newblock PLoS one. 2010;5(2):e9012.

\bibitem{nowak2011demographic}
Nowak K, Lee PC.
\newblock Demographic structure of Zanzibar red colobus populations in unprotected coral rag and mangrove forests.
\newblock International Journal of Primatology. 2011;32(1):24-45.
\newblock doi:10.1007/s10764-010-9434-2.

\bibitem{snaith2008red}
Snaith TV, Chapman CA.
\newblock Red colobus monkeys display alternative behavioral responses to the costs of scramble competition.
\newblock Behavioral Ecology. 2008;19(6):1289-96.
\newblock doi:10.1093/beheco/arn076.

\bibitem{angulo2013social}
Angulo E, Rasmussen GS, Macdonald DW, Courchamp F.
\newblock Do social groups prevent {A}llee effect related extinctions?: The case of wild dogs.
\newblock Frontiers in zoology. 2013;10(1):1-14.

\bibitem{luque2013allee}
Luque GM, Giraud T, Courchamp F.
\newblock Allee effects in ants.
\newblock Journal of Animal Ecology. 2013;82(5):956-65.
\newblock doi:10.1111/1365-2656.12091.

\bibitem{angulo2018allee}
Angulo E, Luque GM, Gregory SD, Wenzel JW, Bessa-Gomes C, Berec L, et~al.
\newblock Allee effects in social species.
\newblock Journal of Animal Ecology. 2018;87(1):47-58.

\bibitem{dennis1989allee}
Dennis B.
\newblock {A}llee effects: population growth, critical density, and the chance of extinction.
\newblock Natural Resource Modeling. 1989;3(4):481-538.

\bibitem{liermann2001depensation}
Liermann M, Hilborn R.
\newblock Depensation: evidence, models and implications.
\newblock Fish and Fisheries. 2001;2(1):33-58.
\newblock doi:10.1046/j.1467-2979.2001.00029.x.

\bibitem{tcheslavskaia2002mating}
Tcheslavskaia K, Brewster CC, Sharov AA.
\newblock Mating success of gypsy moth (Lepidoptera: Lymantriidae) females in southern Wisconsin.
\newblock The Great Lakes Entomologist. 2002;35(1):1.
\newblock doi:10.22543/0090-0222.2054.

\bibitem{garrett2002allee}
Garrett K, Bowden R.
\newblock An {A}llee effect reduces the invasive potential of Tilletia indica.
\newblock Phytopathology. 2002;92(11):1152-9.

\bibitem{ashman2004pollen}
Ashman TL, Knight TM, Steets JA, Amarasekare P, Burd M, Campbell DR, et~al.
\newblock Pollen limitation of plant reproduction: ecological and evolutionary causes and consequences.
\newblock Ecology. 2004;85(9):2408-21.
\newblock doi:10.1890/03-8024.

\bibitem{wagenius2006scale}
Wagenius S.
\newblock Scale dependence of reproductive failure in fragmented Echinacea populations.
\newblock Ecology. 2006;87(4):931-41.
\newblock doi:10.1890/0012-9658(2006)87%5B931:SDORFI%5D2.0.CO;2.

\bibitem{luzuriaga2006population}
Luzuriaga AL, Escudero A, Albert MJ, Gim{\'e}nez-Benavides L.
\newblock Population structure effect on reproduction of a rare plant: beyond population size effect.
\newblock Botany. 2006;84(9):1371-9.
\newblock doi:10.1139/b06-078.

\bibitem{lundquist2011estimating}
Lundquist CJ, Botsford LW.
\newblock Estimating larval production of a broadcast spawner: the influence of density, aggregation, and the fertilization {A}llee effect.
\newblock Canadian Journal of Fisheries and Aquatic Sciences. 2011;68(1):30-42.
\newblock doi:10.1139/F10-125.

\bibitem{guy2019importance}
Guy C, Smyth D, Roberts D.
\newblock The importance of population density and inter-individual distance in conserving the European oyster Ostrea edulis.
\newblock Journal of the Marine Biological Association of the United Kingdom. 2019;99(3):587-93.

\bibitem{Morris2002}
Morris DW.
\newblock {Measuring the Allee effect: positive density dependence in small mammals}.
\newblock Ecology. 2002;83(1):14-20.
\newblock doi:10.1890/0012-9658(2002)083%5B0014:MTAEPD%5D2.0.CO;2.

\bibitem{Cassini2011}
Cassini MH.
\newblock {Consequences of local Allee effects in spatially structured populations}.
\newblock Oecologia. 2011;165(3):547-52.
\newblock doi:10.1007/s00442-010-1897-5.

\bibitem{kramer2009evidence}
Kramer AM, Dennis B, Liebhold AM, Drake JM.
\newblock The evidence for {A}llee effects.
\newblock Population Ecology. 2009;51(3):341-54.

\bibitem{kanarek2013allee}
Kanarek AR, Webb CT, Barfield M, Holt RD.
\newblock Allee effects, aggregation, and invasion success.
\newblock Theoretical ecology. 2013;6(2):153-64.
\newblock doi:10.1007/s12080-012-0167-z.

\bibitem{Surendran2020}
Surendran A, Plank MJ, Simpson MJ.
\newblock {Population dynamics with spatial structure and an {A}llee effect}.
\newblock Proceedings of the Royal Society A: Mathematical, Physical and Engineering Sciences. 2020;476:1-19.

\bibitem{allee1938social}
Allee WC.
\newblock Social life of animals.
\newblock Edn 1. William Heineman Ltd, London and Toronto; 1938.

\bibitem{le2008allee}
Le~Cadre S, Tully T, Mazer SJ, Ferdy JB, Moret J, Machon N.
\newblock Allee effects within small populations of \textit{Aconitum napellus ssp. lusitanicum}, a protected subspecies in {N}orthern {F}rance.
\newblock New Phytologist. 2008;179(4):1171-82.

\bibitem{silliman2015facilitation}
Silliman BR, Schrack E, He Q, Cope R, Santoni A, van~der Heide T, et~al.
\newblock Facilitation shifts paradigms and can amplify coastal restoration efforts.
\newblock Proceedings of the National Academy of Sciences. 2015;112(46):14295-300.
\newblock doi:10.1073/pnas.1515297112.

\bibitem{lerch2018demographic}
Lerch BA, Nolting BC, Abbott KC.
\newblock Why are demographic {A}llee effects so rarely seen in social animals?
\newblock Journal of Animal Ecology. 2018;87(6):1547-59.

\bibitem{woodroffe2020within}
Woodroffe R, O'Neill HM, Rabaiotti D.
\newblock Within-and between-group dynamics in an obligate cooperative breeder.
\newblock Journal of Animal Ecology. 2020;89(2):530-40.
\newblock doi:10.1111/1365-2656.13102.

\bibitem{volterra1938}
Volterra V.
\newblock Population growth, equilibria, and extinction under specified breeding conditions: a development and extension of the theory of the logistic curve.
\newblock Human Biology. 1938;10(1):1-11.
\newblock doi:10.1007/978-3-642-50151-7\_3.

\bibitem{kostitzin1940}
Kostitzin V.
\newblock Sur la loi logistique et ses g{\'e}n{\'e}ralisations.
\newblock Acta Biotheoretica. 1940;5(3):155-9.

\bibitem{tammes1964sexual}
Tammes P, Klomp H, Van~Montfort M.
\newblock Sexual reproduction and underpopulation.
\newblock Archives N{\'e}erlandaises de Zoologie. 1964;16(1):105-10.
\newblock doi:10.1163/036551664X00022.

\bibitem{hsu1975population}
Hsu PH, Fredrickson A.
\newblock Population-changing processes and the dynamics of sexual populations.
\newblock Mathematical Biosciences. 1975;26(1-2):55-78.
\newblock doi:10.1016/0025-5564(75)90094-2.

\bibitem{asmussen1979density}
Asmussen MA.
\newblock Density-dependent selection II. The {A}llee effect.
\newblock The American Naturalist. 1979;114(6):796-809.

\bibitem{cushing1988}
Cushing J.
\newblock 1.
\newblock In: Hallam TG, Gross L, Levin S, editors. The {A}llee effect in age-structured population dynamics. World Scientific Publ.; 1988. p. 479-505.

\bibitem{dennis1981extinction}
Dennis B.
\newblock Extinction and waiting times in birth-death processes: applications to endangered species and insect pest control.
\newblock Statistical distributions in scientific work. 1981;6:289-301.
\newblock doi:10.1007/978-94-009-8555-1\_19.

\bibitem{dennis2002}
Dennis B.
\newblock Allee effects in stochastic populations.
\newblock Oikos. 2002;96(3):389-401.
\newblock doi:10.1034/j.1600-0706.2002.960301.x.

\bibitem{mendez2019demographic}
M{\'e}ndez V, Assaf M, Mas{\'o}-Puigdellosas A, Campos D, Horsthemke W.
\newblock Demographic stochasticity and extinction in populations with {A}llee effect.
\newblock Physical Review E. 2019;99(2):022101.

\bibitem{rijnsdorp2001feeding}
Rijnsdorp A, Vingerhoed Bv.
\newblock Feeding of plaice \textit{Pleuronectes platessa} L. and sole \textit{Solea solea} (L.) in relation to the effects of bottom trawling.
\newblock Journal of Sea Research. 2001;45(3-4):219-29.
\newblock doi:10.1016/S1385-1101(01)00047-8.

\bibitem{padron2000effect}
Padr{\'o}n V, Trevisan MC.
\newblock Effect of aggregating behavior on population recovery on a set of habitat islands.
\newblock Mathematical biosciences. 2000;165(1):63-78.
\newblock doi:10.1016/S0025-5564(00)00005-5.

\bibitem{Fadai2019}
Fadai NT, Johnston ST, Simpson MJ.
\newblock Unpacking the {A}llee effect: determining individual-level mechanisms that drive global population dynamics.
\newblock Proceedings of the Royal Society A. 2020;476(2241):20200350.

\bibitem{keitt2001allee}
Keitt TH, Lewis MA, Holt RD.
\newblock Allee effects, invasion pinning, and species’ borders.
\newblock The American Naturalist. 2001;157(2):203-16.

\bibitem{maciel2015allee}
Maciel GA, Lutscher F.
\newblock Allee effects and population spread in patchy landscapes.
\newblock Journal of Biological Dynamics. 2015;9(1):109-23.
\newblock doi:10.1080/17513758.2015.1027309.

\bibitem{takasu2009}
Takasu F.
\newblock Individual-based modeling of the spread of pine wilt disease: vector beetle dispersal and the Allee effect.
\newblock Population Ecology. 2009;51:399-409.
\newblock doi:10.1007/s10144-009-0145-5.

\bibitem{berec2001}
Berec L, Boukal DS, Berec M.
\newblock Linking the Allee effect, sexual reproduction, and temperature-dependent sex determination via spatial dynamics.
\newblock The American Naturalist. 2001;157(2):217-30.
\newblock doi:10.1086/318626.

\bibitem{boukal2002}
Boukal DS, Berec L.
\newblock Single-species models of the Allee effect: extinction boundaries, sex ratios and mate encounters.
\newblock Journal of Theoretical Biology. 2002;218(3):375-94.
\newblock doi:10.1006/jtbi.2002.3084.

\bibitem{hernandez2004clustering}
Hern{\'a}ndez-Garc{\'\i}a E, L{\'o}pez C.
\newblock Clustering, advection, and patterns in a model of population dynamics with neighborhood-dependent rates.
\newblock Physical Review E. 2004;70(1):016216.

\bibitem{Crews1986}
Crews D, Grassman M, Lindzey J.
\newblock {Behavioral facilitation of reproduction in sexual and unisexual whiptail lizards.}
\newblock Proceedings of the National Academy of Sciences of the United States of America. 1986;83(24):9547-50.
\newblock doi:10.1073/pnas.83.24.9547.

\bibitem{Thomas1974}
Thomas JD, Benjamin M.
\newblock {The Effects of Population Density on Growth and Reproduction of Biomphalaria glabrata (Say) (Gasteropoda: Pulmonata)}.
\newblock The Journal of Animal Ecology. 1974;43(1):31.
\newblock doi:10.2307/3156.

\bibitem{doering2003}
Doering CR, Mueller C, Smereka P.
\newblock Interacting particles, the stochastic Fisher--Kolmogorov--Petrovsky--Piscounov equation, and duality.
\newblock Physica A: Statistical Mechanics and its Applications. 2003;325(1-2):243-59.
\newblock doi:10.1016/S0378-4371(03)00203-6.

\bibitem{toral2014stochastic}
Toral R, Colet P.
\newblock Stochastic numerical methods: an introduction for students and scientists.
\newblock John Wiley \& Sons; 2014.
\newblock doi:10.1002/9783527683147.

\bibitem{constable2016}
Constable G, Rogers T, McKane AJ, Tarnita CE.
\newblock {Demographic noise can reverse the direction of deterministic selection}.
\newblock Proceedings of the National Academy of Sciences USA. 2016:4745-54.
\newblock doi:10.1073/pnas.1603693113.

\bibitem{oro2020}
Oro D.
\newblock Perturbation, behavioural feedbacks, and population dynamics in social animals: when to leave and where to go.
\newblock Oxford University Press, USA; 2020.
\newblock doi:10.1093/oso/9780198849834.001.0001.

\bibitem{nisbet1982}
Nisbet RM, Gurney W.
\newblock 6.
\newblock In: Birth and death processes in a static environment. Blackburn Press; 1982. p. 167-220.

\bibitem{renshaw1993}
Renshaw E.
\newblock 3.
\newblock In: General birth-death processes. Cambridge University Press; 1993. p. 46-86.
\newblock doi:10.1017/CBO9780511624094.005.

\bibitem{allee1949principles}
Allee WC, Park O, Emerson AE, Park T, Schmidt KP, et~al.
\newblock Principles of {A}nimal {E}cology.
\newblock Edn 1. WB Saundere Co. Ltd.; 1949.

\bibitem{doi1976}
Doi M.
\newblock Stochastic theory of diffusion-controlled reaction.
\newblock Journal of Physics A: Mathematical and General. 1976;9(9):1479.
\newblock doi:10.1088/0305-4470/9/9/009.

\bibitem{peliti1985}
Peliti L.
\newblock Path integral approach to birth-death processes on a lattice.
\newblock Journal de Physique. 1985;46(9):1469-83.
\newblock doi:10.1051/jphys:019850046090146900.

\bibitem{Tauber2007}
T{\"a}uber UC.
\newblock 7.
\newblock In: Field-Theory Approaches to Nonequilibrium Dynamics. Berlin, Heidelberg: Springer Berlin Heidelberg; 2007. p. 295-348.
\newblock Available from: \url{https://doi.org/10.1007/3-540-69684-9_7}. doi:10.1007/3-540-69684-9\_7.

\bibitem{Cross1993}
Cross MC, Hohenberg PC.
\newblock {Pattern formation outside of equilibrium}.
\newblock Reviews of Modern Physics. 1993;65(3).
\newblock Available from: \url{http://rmp.aps.org/abstract/RMP/v65/i3/p851_1}. doi:10.1103/RevModPhys.65.851.

\bibitem{kot2001ch2}
Kot M.
\newblock 2.
\newblock In: Harvest models: bifurcations and breakpoints. Cambridge University Press; 2001. p. 13-25.
\newblock doi:10.1017/CBO9780511608520.003.

\bibitem{Gillespie1977}
Gillespie DT.
\newblock {Exact stochastic simulation of coupled chemical reactions}.
\newblock The journal of physical chemistry. 1977;93555(1):2340-61.
\newblock Available from: \url{http://pubs.acs.org/doi/abs/10.1021/j100540a008}. doi:10.1021/j100540a008.

\bibitem{Martinez-Garcia2015}
Martinez-Garcia R, Murgui C, Hern{\'{a}}ndez-Garc{\'{i}}a E, L{\'{o}}pez C.
\newblock {Pattern Formation in Populations with Density-Dependent Movement and Two Interaction Scales}.
\newblock PLoS ONE. 2015;10:e0132261.
\newblock doi:10.1371/journal.pone.0132261.

\bibitem{rietkerk2008regular}
Rietkerk M, Van~de Koppel J.
\newblock Regular pattern formation in real ecosystems.
\newblock Trends in ecology \& evolution. 2008;23(3):169-75.
\newblock doi:10.1016/j.tree.2007.10.013.

\bibitem{martinez2013vegetation}
Mart{\'\i}nez-Garc{\'\i}a R, Calabrese JM, Hern{\'a}ndez-Garc{\'\i}a E, L{\'o}pez C.
\newblock Vegetation pattern formation in semiarid systems without facilitative mechanisms.
\newblock Geophysical Research Letters. 2013;40(23):6143-7.

\bibitem{martinez2014minimal}
Mart{\'\i}nez-Garc{\'\i}a R, Calabrese JM, Hern{\'a}ndez-Garc{\'\i}a E, L{\'o}pez C.
\newblock Minimal mechanisms for vegetation patterns in semiarid regions.
\newblock Philosophical Transactions of the Royal Society A: Mathematical, Physical and Engineering Sciences. 2014;372(2027):20140068.
\newblock doi:10.1098/rsta.2014.0068.

\bibitem{martinezgarcia2022}
Martinez-Garcia R, Cabal C, Calabrese JM, Hern{\'a}ndez-Garc{\'\i}a E, Tarnita CE, L{\'o}pez C, et~al.
\newblock Integrating theory and experiments to link local mechanisms and ecosystem-level consequences of vegetation patterns in drylands.
\newblock Chaos, Solitons \& Fractals. 2023;166:112881.
\newblock doi:10.1016/j.chaos.2022.112881.

\bibitem{kao2023}
Kao AB, Hund AK, Santos FP, Young JG, Bhat D, Garland J, et~al.
\newblock {Opposing Responses to Scarcity Emerge from Functionally Unique Sociality Drivers}.
\newblock American Naturalist. 2023;202(3):302-21.
\newblock doi:10.1086/725426.

\bibitem{Fronhofer2015}
Fronhofer EA, Kropf T, Altermatt F.
\newblock Density-dependent movement and the consequences of the Allee effect in the model organism Tetrahymena.
\newblock Journal of Animal Ecology. 2015;84(3):712-22.
\newblock doi:10.1111/1365-2656.12315.

\bibitem{woodroffe2011demography}
Woodroffe R.
\newblock Demography of a recovering African wild dog (Lycaon pictus) population.
\newblock Journal of Mammalogy. 2011;92(2):305-15.
\newblock doi:10.1644/10-MAMM-A-157.1.

\bibitem{lewis1993allee}
Lewis MA, Kareiva P.
\newblock Allee dynamics and the spread of invading organisms.
\newblock Theoretical Population Biology. 1993;43(2):141-58.
\newblock doi:10.1006/tpbi.1993.1007.

\bibitem{Heinsalu2013}
Heinsalu E, Hern{\'a}ndez-Garcia E, L{\'o}pez C.
\newblock Clustering determines who survives for competing brownian and l{\'e}vy walkers.
\newblock Physical Review Letters. 2013;110(25):258101.

\bibitem{Pigolotti2014}
Pigolotti S, Benzi R.
\newblock Selective advantage of diffusing faster.
\newblock Physical Review Letters. 2014;112(18):188102.
\newblock doi:10.1103/PhysRevLett.112.188102.

\bibitem{Shaw2015}
Shaw AK, Kokko H.
\newblock Dispersal evolution in the presence of Allee effects can speed up or slow down invasions.
\newblock The American Naturalist. 2015;185(5):631-9.
\newblock doi:10.1086/680511.

\bibitem{pigolotti2007}
Pigolotti S, L{\'o}pez C, Hern{\'a}ndez-Garc{\'\i}a E.
\newblock Species clustering in competitive Lotka-Volterra models.
\newblock Physical review letters. 2007;98(25):258101.
\newblock doi:10.1103/PhysRevLett.98.258101.

\bibitem{colombo2022}
Colombo EH, L{\'o}pez C, Hern{\'a}ndez-Garc{\'\i}a E.
\newblock Pulsed interaction signals as a route to biological pattern formation.
\newblock Physical Review Letters. 2023;130(5):058401.
\newblock doi:10.1103/PhysRevLett.130.058401.

\bibitem{colombo2023pulsed}
Colombo EH, Martinez-Garcia R, Calabrese JM, L{\'o}pez C, Hern{\'a}ndez-Garc{\'\i}a E.
\newblock Pulsed interactions unify reaction-diffusion and spatial nonlocal models for biological pattern formation.
\newblock arXiv preprint arXiv:231107135. 2023.

\bibitem{liu2013}
Liu QX, Doelman A, Rottsch{\"a}fer V, de~Jager M, Herman PM, Rietkerk M, et~al.
\newblock Phase separation explains a new class of self-organized spatial patterns in ecological systems.
\newblock Proceedings of the National Academy of Sciences. 2013;110(29):11905-10.
\newblock doi:10.1073/pnas.1222339110.

\bibitem{rao2016}
Rao F, Kang Y.
\newblock The complex dynamics of a diffusive prey--predator model with an {A}llee effect in prey.
\newblock Ecological complexity. 2016;28:123-44.

\bibitem{Martinez-Garcia2022}
Martinez-Garcia R, Tarnita CE, Bonachela JA.
\newblock {Self-organized patterns in ecological systems: from microbial colonies to landscapes}.
\newblock Emerging Topics in Life Sciences. 2022;6(3):245-58.
\newblock doi:10.1042/ETLS20210282.

\bibitem{maciel2021}
Maciel GA, Martinez-Garcia R.
\newblock Enhanced species coexistence in {L}otka-{V}olterra competition models due to nonlocal interactions.
\newblock Journal of Theoretical Biology. 2021;530:110872.

\bibitem{simoy2023}
Simoy MI, Kuperman MN.
\newblock Non-local interaction effects in models of interacting populations.
\newblock Chaos, Solitons \& Fractals. 2023;167:112993.
\newblock doi:10.1016/j.chaos.2022.112993.

\bibitem{zenodo}
Jorge D. Code for: Demographic effects of aggregation in the presence of a component Allee effect; 2024.
\newblock Available from: \url{https://zenodo.org/records/10790944}. doi:10.5281/zenodo.10790944.

\end{thebibliography}

\end{document}

\maketitle
%\vspace{-0.2cm}
%\setcounter{tocdepth}{1}
\tableofcontents
\newpage
\section{Derivation of the population-level approximation Eq.\,(2.8)}\label{sec:appMF}

In this section, we detail the steps to obtain a deterministic equation for the dynamics of the density of individuals starting from the stochastic individual-level reactions. To this end, we apply the Doi-Peliti formalism, which is a field-theoretical approach, developed in the context of statistical field theory, that uses path integrals to map an individual-level stochastic dynamics to a continuum description in terms of a density field \citep{doi1976,peliti1985,Tauber2007,hernandez2004clustering}.

\subsection{Derivation of the Master equation}\label{sec:AppMEQ}

The Master equation characterizes the evolution of the probability that a system following a stochastic dynamics is in a specific state at a given time $t$. In our case, the Master equation describes how the probability of finding a certain population size and spatial distribution of individuals across the lattice nodes changes with time. We denote this lattice configuration by a vector $\eta$ that specifies the number of individuals in each lattice node: $\eta=\{...n_{i-1},\,n_{i},\,n_{i+1}...\}$. Each lattice node coordinate, labeled by the index $i$, can be mapped to a spatial coordinate $x_i$ using the transformation $x_i=i\,\delta x$, where $\delta x$ is the distance between two adjacent lattice nodes. 

To construct the Master equation, we need to obtain the global transition rates, $\Omega(\eta\to\eta')$, that define the probabilistic transitions between two lattice configurations $\eta$ and $\eta'$ and will depend on the birth, death, and movement stochastic events. Using these global transition rates, we can write the Master equation as
\begin{equation}\label{mastergen}
     \dfrac{\partial P(\eta,t)}{\partial t}= \sum_{\eta'} \Omega(\eta'\to \eta)P(\eta',t) - \Omega(\eta\to\eta')P(\eta,t)
\end{equation}

\subsubsection*{Contribution of birth processes to the global transition rates}

Birth processes contribute to the appearance of a new individual in a focal lattice position $x$ via density-independent reproduction and facilitation. These processes are represented by the following biological reactions
\begin{align}
    &\emptycirc[0.9ex]_x \xrightarrow{b} \emptycirc[0.9ex]_x + \emptycirc[0.9ex]_x ,\label{eq: birth}\\
    &\emptycirc[0.9ex]_x +  \emptycirc[0.9ex]_{x'} \xrightarrow{\frac{\beta}{2 R_f}} \emptycirc[0.9ex]_x + \emptycirc[0.9ex]_{x'} + \emptycirc[0.9ex]_{x}, \label{eq:faci}
\end{align}
where the reaction \eqref{eq:faci} only takes place if $\rvert x-x'\rvert \leq R_f$. We can decompose the global transition rate resulting from these birth processes in two demographic rates $W$, one corresponding to each of the reactions that can potentially contribute to the transition from a configuration $\eta$ to $\eta'=\{ n_x+1\}_\eta$:
\begin{equation}
\Omega\big(\eta\to\{ n_x+1\}_\eta\big)= W_b(n_x)+W_{\beta}(\eta,x), \label{eq:omegab}
\end{equation}
where $\{n_x+1\}_\eta$ denotes a lattice configuration in which all nodes have the same number of individuals as in the configuration $\eta$ except the node with spatial coordinate $x$, where the occupancy has increased in one unity. $W_b(n_x)$ and $W_\beta(\eta,x)$ are the demographic rates in which the reactions for density-independent birth and facilitation, \eqref{eq: birth} and \eqref{eq:faci} respectively, generate individuals at the position $x$.

The demographic rate corresponding to density-independent birth increases linearly with the number of individuals in the position $x$
\begin{equation}
     W_b(n_x)=b \,n_x \label{rb}.
\end{equation}
For the facilitation demographic rate, however, we need to take into account the long-range of the interaction and thus the fact that reproduction at the lattice coordinate $x$ depends on the number of individuals within the lattice nodes such that $\rvert x-x'\rvert\leq R_f$, $N^{f}_x$. For example, the demographic rate in which the facilitation reaction \eqref{eq:faci} generates an individual in $x$ changes depending on whether $x'$ is equal or different from $x$. This difference exists because pairwise facilitation only increases the number of individuals at $x$ half of the times if $x'\neq x$, leading to a new individual at $x'$ the other half. For $x=x'$, however, the new individual is always located at $x$. Thus, considering both the number of pairs we can form for $x=x'$ and $x\neq x'$, the facilitation demographic rate is
\begin{equation}
    W_\beta(\eta,x) = \dfrac{\beta}{2R_f}\Bigg[\overbrace{ \begin{pmatrix} n_x\\
2
\end{pmatrix}}^{x'=x} + \underbrace{\dfrac{1}{2}(N^{f}_x-n_x)n_x}_{x'\neq x}\Bigg]
\end{equation}
which simplifies to
\begin{equation}
    W_\beta(\eta,x) = \dfrac{\beta \, n_x}{4 \, R_f}\left(N^{f}_x-1\right)\label{rbt}.
\end{equation}

\subsubsection*{Contribution of death processes to the demographic transition rates}

Individuals die in each lattice coordinate $x$ due to density-independent death and competition, which we can write in terms of biological reactions as
\begin{align}
    &\emptycirc[0.9ex]_x \xrightarrow{d} \text{\O},\label{eq: death}\\
    &\emptycirc[0.9ex]_x + \emptycirc[0.9ex]_{x'} + \emptycirc[0.9ex]_{x''} \xrightarrow{\frac{\gamma}{4 R_c^2}} \emptycirc[0.9ex]_{x'} + \emptycirc[0.9ex]_{x''}, \label{eq:comp}
\end{align}
\eqref{eq:comp} only takes place if both pairwise distances $\rvert x-x'\rvert\leq R_c$ and $\rvert x-x''\rvert\leq R_c$. The global transition rate due to death processes is 
\begin{equation}
\Omega(\eta\to\{ n_x-1\}_\eta)= W_d(n_x)+W_{\gamma}(\eta,x) \label{eq:omegad}
\end{equation}
where $W_d(n_x)$ and $W_\gamma(\eta,x)$ are the demographic rates in which the density-independent death and competition reactions, \eqref{eq: death} and \eqref{eq:comp} respectively, eliminate an individual at lattice coordinate $x$.

Similarly to the density-independent birth process, the density-independent death demographic rate increases linearly with $n_x$
\begin{eqnarray}\label{}
     W_d(n_x)=d\,n_x \label{rd},
\end{eqnarray} 
and extending the same combinatorial arguments used to derive Eq.\,\eqref{rbt} we obtain the ternary competition demographic rate that results in the death of an individual at lattice coordinate $x$, 
\begin{equation}
    W_\gamma(\eta,x) =\dfrac{\gamma}{6} \dfrac{n_x}{4R_c^2}\left(N^{c}_x-1\right)\left(N^{c}_x-2\right)\label{rg}
\end{equation}
where $N^{c}_x$ is the number of organisms within a distance $R_c$ from $x$.

\subsubsection*{Contribution of movement to the demographic transition rates}
The stochastic event that leads to individual movement is
\begin{eqnarray} \label{eq:jumps}
    \emptycirc[0.9ex]_x \xrightarrow{h} \emptycirc[0.9ex]_{x\pm \delta x}.
\end{eqnarray}
where the subscript indicates the location of the individual in spatial coordinates. Because this is a density-independent process, the global rate is linear in $n_x$ and is given by
\begin{equation}
    \Omega(\eta\to\{ n_x-1,n_{x'}+1\}_\eta)=  h n_x\label{movementd}.
\end{equation}
where $x'=x+\delta x$ is a nearest neighbor of $x$.

\vspace{1cm}

\noindent Finally, combining Eqs.\,\eqref{eq:omegab}, \eqref{eq:omegad}, and \eqref{rd} for the global transition rates and scaling $\beta\to2\beta$ and $\gamma\to2\gamma$, we can obtain the explicit form of the Master equation. Here, we map the positions to their lattice site counterparts $x \mapsto i$.

\begin{eqnarray}\label{eq:master}
          \dfrac{\partial P(\eta,t)}{\partial t}&=& b\sum_i  (n_i-1)P\left(\{n_i-1\}_\eta,t\right) -  n_i P(\eta,t)
          +d\sum_i  (n_i+1) P\left(\{n_i+1\}_\eta,t\right) -  n_i P(\eta,t)\nonumber\\
          &&+\dfrac{\beta}{2R_f}\sum_i (n_i-1) (N_i^f - 2) P\left(\{n_i-1\}_\eta,t\right) -  n_i (N_i^f -1) P(\eta,t)\nonumber\\
          &&+\dfrac{\gamma}{4R^2_c}\sum_i (n_i+1) N_i^c (N_i^c-1)P\left(\{n_i+1\}_\eta,t\right)- n_i (N_i^c-1)(N_i^c-2)P\left(\eta,t\right)\nonumber\\
          &&+ h\sum_{\langle ij\rangle}(n_i +1) P\left(\{n_i +1,\,n_j-1\}_\eta,t\right) - n_i P(\eta, t).
\end{eqnarray}
where $n_i$ is the number of individuals at the lattice site $i$ and $N_{i}^\alpha$ denotes the number of individuals within the range $R_\alpha$ centered at $i$. The notation $\langle ij\rangle$ specifies that the sum is performed over the nearest neighbors of $i$.

\subsection{Description of the population configuration in terms of a Fock space}
Individuals in our discrete model are represented by indistinguishable particles distributed within the cells of a one-dimensional lattice. Therefore, as introduced in Section \ref{sec:AppMEQ}, we can describe the state of the system at a given time if we know the number of individuals (particles) in each lattice cell (lattice configuration), $\eta=\{n_0...n_i,\,n_{i+1},\,n_{i+2}...n_N\}$. To apply the Doi-Peliti formalism, we first need to define a Fock space for each lattice node with a basis given by the occupancy number basis and their corresponding creation and annihilation operators. These operators will be responsible for updating the state of each lattice node by changing the number of particles they contain. Therefore, these creation and annihilation operators are related to the demographic rates in our model. Using Dirac notation, the occupancy number basis is a set of vectors $\ket{n_i}$, giving the number of individuals, $n$, in each lattice cell $i$. The creation and annihilation operators, $a^\dagger_i$ and $a_i$, respectively, act on each vector of this basis following
\begin{align}
&a_{i}^{\dagger}\left|n_{i}\right\rangle=\left|n_{i}+1\right\rangle \label{eq:a} \\
&a_{i}\left|n_{i}\right\rangle=n_{i}\left|n_{i}-1\right\rangle \label{eq:a2}
\end{align}
and have bosonic commutation relations given by
\begin{align}
&{\left[a_{i}, a_{j}^{\dagger}\right]=a_{i} a_{j}^{\dagger}-a_{j}^{\dagger} a_{i}=\delta_{i j}} \label{eq:r} \\
&{\left[a_{i}^{\dagger}, a_{j}^{\dagger}\right]=\left[a_{i}, a_{j}\right]=0}. \label{eq:r2}
\end{align}

Finally, using the occupancy number basis for each lattice node and Dirac's notation, we can express the basis associated with the lattice configuration $\eta$ as the tensor product of the basis of each site
\begin{align}
    \ket{\eta} = \ket{n_0}\otimes...\ket{n_{i-1}} \otimes \ket{n_i} \otimes \ket{n_{i+1}} \otimes \hdots \ket{n_N}.
\end{align}

\subsection{Derivation of a dynamical equation for population configurations}
The dynamics of our individual-based model is stochastic. Hence, to fully characterize its dynamics, we must describe how the probability that the system is observed in a given configuration, $\eta$, changes with time. To this end, we first define a general lattice configuration at a time $t$, $\ket{\psi(t)}$, as the sum of all of the possible lattice configurations, i.e.\,the basis states, weighted by the probability of observing each of them
\begin{equation}\label{State}
\ket{\psi(t)}=\sum_{\eta}P(\eta,t)\,\ket{\eta},
\end{equation}
which changes in time according to
\begin{align}\label{ev}
    \dfrac{\partial}{\partial t}\ket{\psi(t)}=\sum_{\eta} \dfrac{\partial}{\partial t}P(\eta,t)\,\ket{\eta}.
\end{align}
Therefore, the dynamics of the state $\ket{\psi(t)}$ is determined by the dynamics of the probabilities of finding the system in each of its possible states. These possible states are described by the basis states $\ket{\eta}$ and the probabilities of finding the system in each of them change according to the Master equation we derived in Section \ref{sec:AppMEQ}. Using some algebra and the commutation relationships for the bosonic operators in Eqs.\,\eqref{eq:a}-\eqref{eq:r2}, we can represent \eqref{ev} as a Schrodinger-like equation 
\begin{align}
     \dfrac{\partial}{\partial t}\ket{\psi(t)}= -H(\{a^\dagger,a\})\ket{\psi(t)}
\end{align}
with a quasi-hamiltonian given by $H(\{a^\dagger,a\})=\sum_i \mathcal{H}_i(\{a^\dagger,a\})$ where
\begin{align}
   \mathcal{H}_i(\{a^\dagger,a\})=&\ b \left(\mathbb{I}-a_i ^\dagger \right)a_i ^\dagger a_i
   \,+\, d  \left(a_i ^\dagger-\mathbb{I} \right) a_i +\frac{\beta}{2R_f}  \left(\mathbb{I}- a^\dagger_i\right) a^\dagger_i \sum_{j \,\in R_f(i)} \left\{ a^\dagger_j a_j\right\} a_i\\
    &+\frac{\gamma}{4R^2_c}  \left(a^\dagger _i-\mathbb{I} \right) \sum_{j,k \,\in R_c(i)} \left\{ a^\dagger_j a^\dagger_k a_k a_j\right\}  a_i
    +\,\dfrac{h}{2} \sum_{\langle ij\rangle} \left(a_i^\dagger -a_j^\dagger \right) \left(a_i  - a_j \right).
\end{align}

%An important property of this quasi-hamiltonian is that it is normally ordered, i.e. creation operators on the left and annihilation operators on the right. 

\subsection{Derivation of a path integral representation in terms of continuous field}
To obtain a continuous equation for the dynamics of a population density field, we next need to derive a path integral representation of our model dynamics. First, we introduce the fields $\phi_i(t)$ and $\phi^*_i(t)$ in the Hamiltonian $\mathcal{H}( a^\dagger \to \phi(t)^*+1,\ a\to \phi(t))$ and then take the continuum limit by letting the lattice spacing $\delta x\to 0$. To this end, we redefine the fields and parameters as
\begin{equation}
    \frac{\phi_{i}(t)}{\delta x} \mapsto \phi(x, t), \quad \phi_i^{*}(t) \mapsto \tilde{\phi}(x, t), \quad \dfrac{h}{2} \mapsto \frac{D}{\delta x^{2}},
\end{equation}
and the quasi-hamiltonian as $H\left(\phi^*(x,t),\phi(x,t)\right)=\int \mathcal{H}\left(\phi^*(x,t),\phi(x,t)\right)dx$. The new fields are associated with an action $S$, given by
\begin{align}\label{S}
S[\phi^*(x, t), \phi(x, t)] &=\int d x \int_{0}^{t} d \tau \, \, \, \phi^*\partial_\tau \phi+ \mathcal{H}[\phi^*(x, \tau), \phi(x, \tau)].
\end{align}

Next, we obtain the dynamic equation of the fields using the stationary-action principle or principle of least action, i.e. $\frac{\delta S}{\delta \phi}=\frac{\delta S}{\delta \tilde{\phi}}=0$. By doing so, we obtain that the terms in the action that are linear in $\phi^*$ give rise to the equation
\begin{equation}
    \dfrac{\partial \phi(x,t)}{\partial t} =\left[ r  + \frac{\beta}{2R_f} \, \int_{x-R_f}^{x+R_f} \phi(x^\prime,t)dx^\prime  - \frac{\gamma}{4R^2_c} \left(\int_{x-R_c}^{x+R_c} \phi(x^\prime,t)dx^\prime \right)^2\right]\phi + D \,\nabla_x^2 \phi.
\end{equation}

Finally, we take the expected value of the field, $\langle \phi \rangle $, which is equivalent to the mean density of particles $\rho$ in the continuum limit. Using the mean-field approximation $\langle \phi^2\rangle \approx \langle \phi\rangle^2=\rho^2$, we obtain
\begin{equation}\label{mf}
    \dfrac{\partial \rho(x,t)}{\partial t} =\left[ r  + \beta \, \tilde{\rho}_f(x,t)   - \gamma \, \tilde{\rho}_c^2(x,t)\right]\rho(x,t) + D \,\nabla_x^2 \rho(x,t).
\end{equation}
where $\nabla_x^2$ is the Laplacian on the spatial coordinate $x$ and $\tilde{\rho}_\alpha(x,t)$ is the non-local density obtained by averaging the population density in the neighborhood of range $R_\alpha$ for $\alpha=\{f,\,c\}$
\begin{equation}\label{trho}
    \tilde{\rho}_\alpha(x, t)=\int G\left(\left|x-x^{\prime}\right|;R_\alpha\right) \rho\left(x^{\prime}, t\right) \mathrm{d} x^{\prime}.
\end{equation}
$G(\left|x-x^{\prime}\right|;R_\alpha)$ is the normalized interaction kernel for the facilitation and competition, $\alpha=\{f,\,c\}$,
\begin{equation}\label{G}
     G(\left|x-x^{\prime}\right|;R_\alpha)= \begin{cases}\frac{1}{2 R_{\alpha}} & \text { if } \left|x-x^{\prime}\right| \leq R_{\alpha} \\ 0 & \text { otherwise. }\end{cases}
\end{equation}

\section{Linear stability analysis of Eq.\,(2.8)}\label{sec:appLSA}

In this section, we provide the detailed steps to perform a linear stability analysis of Eq.~\eqref{mf}, which gives the conditions to observe non-uniform spatial patterns of population density \citep{cross1993pattern}. The linear stability analysis consists in introducing a small perturbation, $\epsilon \, \psi(x,t)$ with $\epsilon \ll 1$, to the uniform stable steady state of a partial differential equation and calculating the time evolution of the perturbation amplitude. If this amplitude decreases with time, the uniform state is also stable against spatial perturbations, and patterns do not form; if the perturbation amplitude increases, the uniform state is unstable against spatial perturbations, and spatial patterns form.

We consider a solution of the form $\rho(x,t) = \rho_+ + \epsilon \, \psi(x,t) $, where $\rho_+$ is the uniform stable state and $\psi$ is the perturbation of amplitude $\epsilon$. Inserting this solution in Eq.\,\eqref{mf} and retaining only linear terms in the perturbation, we get
\begin{equation}\label{psi}
    \dfrac{\partial \psi(x,t)}{\partial t} =  r  \psi(x,t) +   \beta \, \rho_+ \Big[\psi(x,t) + \tilde{\psi}_{f}(x,t)\Big] - \gamma \, \rho_+^2 \Big[ \psi (x,t)+2\tilde{\psi}_{c}(x,t)\Big]    +  D \, \nabla^2 \psi(x,t) 
\end{equation}
where we have introduced the simplifying notation 
\begin{align}
    \tilde{\psi}_{\alpha}(x,t) = \int G_{\alpha} (|x-a|) \,  \psi(x,t) \, da.
\end{align}

Equation \eqref{psi} is a linear integro-differential equation that we can solve using a Fourier transform:
\begin{equation}\label{psih}
    \dfrac{\partial \hat{\psi}(k,t)}{\partial t}= \Big [  r  +   \beta \, \rho_+ \Big(1 + \hat{G}_{f}(k)\Big) -  \gamma \, \rho_+^2 \Big( 1 +2\hat{G}_{c}(k)\Big)    -   D k ^2 \Big] \hat{\psi}(k,t)
\end{equation}
where $\hat{\psi}(k,t)$ is the Fourier transform of the perturbation and $\hat{G}_\alpha(k)$ is the Fourier transform of the kernel
\begin{equation}
    \hat{G}_\alpha(k)= \dfrac{\sin(R_\alpha k)}{R_\alpha\,k}.
\end{equation}

Finally, because Eq.\,\eqref{psih} is a linear differential equation, it has an exponential solution $\hat{\psi}(k,t) \propto \exp{\left(\lambda(k)t\right)}$ whose growth rate $\lambda(k)$ we can obtain from Eq.\,\eqref{psih}
\begin{equation}
    \lambda(k) =  \rho_+ \Big[ \beta \dfrac{\sin(R_f\,k)}{R_f\,k} -2 \gamma \rho_+ \dfrac{\sin(R_c\,k)}{R_c\,k}  \Big ] - D k^2.
\end{equation}

We can also find the nondimensional counterpart of the growth-rate of the perturbation $\overline{\lambda}$, given by 
\begin{equation}
    \overline{\lambda}(\kappa) =  u_+ \Big[  \dfrac{\sin(R\,\kappa)}{R\,\kappa} -2  u_+ \dfrac{\sin(\kappa)}{\kappa}  \Big ] - \overline{D} \kappa^2
\end{equation}
with the nondimensional wavenumber $\kappa=R_c k$.

\section{Numerical computation of the stationary states of Eq.\,(2.8)}\label{sec:appbif}

We compute the stationary states of Eq.\,(2.8) (main text) using its dimensionless counterpart in Eq.\,\eqref{non-dim}. First, we integrate Eq.\,\eqref{non-dim} with $\overline{r}=0$ until the population density reaches its stationary spatial pattern and compute the population abundance by integrating the density field over the system length. Next, we decrease $\overline{r}$ in a small amount $\Delta \overline{r}$ and integrate Eq.\,\eqref{non-dim} for a long time interval $\Delta \tau$ using the stationary pattern for $\overline{r}=0$ as the initial condition. We repeat this process recursively until the $\overline{r}$ is such that the population density vanishes. This procedure gives us the stationary values of the population abundance that are stable, $\overline{u}_{+}$. To compute the unstable ones, we assume that $\overline{u}_{-}(\chi,\tau)= \alpha \, \overline{u}_{+}(\chi,\tau)$, where $\alpha$ is a dimensionless scaling parameter such that $0<\alpha\leq 1$. This assumption implies that $\overline{u}_{-}$ has the same spatial structure as $\overline{u}_{+}$ and that both solutions only differ by a factor $\alpha$ that makes $\overline{u}_{-}(\chi,\tau) < \alpha \,\overline{u}_{+}(\chi,\tau)$. Under this assumption, to compute $\overline{u}_{-}$ we need to numerically find the $0<\alpha\leq1$ that satisfies $\partial_\tau \overline{u}_{-}=0$. Because $\alpha=1$ always satisfies this condition, we only take into account the lowest value of $\alpha$, which is $\alpha=1$ only when $\overline{r}=\overline{r}_c$.

\section{Derivation of the group-level approximation in Eq.\,(3.3) }\label{app:meta}

We build this approximation to obtain estimates for the number of individuals within a single spatial aggregate of the stationary pattern using the three features of the spatial pattern discussed in the main text section 3.2:
\begin{enumerate}
    \item All individuals within a group must interact with one another via competition and facilitation
    \item Individuals of different groups must not compete with each other
    \item If two groups interact with each other via facilitation, this positive interaction must reach all the individuals in both groups
\end{enumerate}

First, we use the definition of $\tilde{\rho}$ in Eq.\,\eqref{trho}. Given that the three assumptions above are fulfilled, $\tilde{\rho}_c$ is constant inside each aggregate. This is so because the integral that defines the averaged densities in Eq.\,\eqref{trho} is equal to the number of individuals within the group, $\mathcal{N}_i$. Therefore,
\begin{equation}\label{1app}
    \tilde{\rho}_c(x_i,t) = \dfrac{\mathcal{N}_I(t)}{2R_c}
\end{equation}
where $x_i$ is the spatial coordinate of each lattice node, $i$, inside the aggregate ``$I$''. Following the same arguments, $\tilde{\rho}_f$ is also constant inside each group. However, since different groups can facilitate each other, the value of $\tilde{\rho}_f$ depends on the number of groups within the facilitation range $R_f$. Thus we can write
\begin{equation}\label{2app}
    \tilde{\rho}_f(x_i,t) = \dfrac{\mathcal{N}_I(t)}{2R_f}+\sum_{\langle I,J\rangle}{\dfrac{\mathcal{N}_J(t)}{2R_f}}
\end{equation}
whereby $J$ runs over the neighbors of the focal group $I$ that are within the facilitation range. Next we use Eqs.\,\eqref{1app} and \eqref{2app} in Eq.\,\eqref{mf} to get:
\begin{equation}\label{3app}
     \dfrac{\partial \rho(x_i,t)}{\partial t} = \Bigg[ r + \beta\dfrac{\mathcal{N}_I(t)}{2R_f}+\beta\sum_{\langle I,J\rangle}{\dfrac{\mathcal{N}_J(t)}{2R_f}} - \gamma \,  \dfrac{\mathcal{N}_I^2(t)}{4 R_c ^2} \Bigg]\rho(x_i,t).
\end{equation}
where we have neglected the diffusion term because small diffusion is a necessary condition to have the three conditions above fulfilled. Finally, we integrate over the group length on both sides of Eq.\,\eqref{3app}
\begin{equation}\label{met_e}
     \dfrac{\partial \mathcal{N}_I(t)}{\partial t} = \Bigg[ r + \dfrac{\beta}{2 R_f}\left(\mathcal{N}_I(t)+\sum_{\langle I,J\rangle} \mathcal{N}_J(t)\right) - \gamma \,  \dfrac{\mathcal{N}_I^2(t)}{4 R_c ^2} \Bigg]\mathcal{N}_I(t).
\end{equation}

If we assume periodic boundary conditions, the spatial pattern of population density is periodic, and all aggregates have the same amount of neighbors and the same aggregate size. Hence, we can introduce a parameter that gives the number of groups within the facilitation range, $\xi$. Using this parameter, we can write an ordinary differential equation to describe the dynamics of aggregate size:
\begin{equation}\label{sm:cubic}
     \dfrac{\partial \mathcal{N}(t)}{\partial t} = \Bigg[ r + \beta  (\xi+1)\, \dfrac{\mathcal{N}(t)}{2 R_f} - \gamma \,  \dfrac{\mathcal{N}^2(t)}{4 R_c ^2} \Bigg]\mathcal{N}(t).
\end{equation}
\noindent Eq.\,\eqref{sm:cubic} is a cubic model with stationary solutions $\mathcal{N}_0=0$ and
\begin{equation}
    \mathcal{N}_{\pm} =   \dfrac{(\xi + 1 )\dfrac{\beta}{2R_f} \pm \sqrt{\Bigg ((\xi + 1 ) \, \dfrac{\beta}{2R_f} \Bigg )^2 +\dfrac{r \gamma}{R_c^2}}}{\gamma/ 2R_c ^2},
\end{equation}
or, in its dimensionless form,
\begin{equation}
    \mathcal{U}_{\pm} =  \dfrac{(\xi + 1 )}{R} \pm \sqrt{ \dfrac{(\xi + 1 )^2}{R^2} +4r}
\end{equation}
where $\mathcal{U}_{\pm}=\gamma\, \mathcal{N}_{\pm}/\beta \, R_c$.

\section{Model nondimensionalization}

To facilitate the model analysis, we define scaled, dimensionless variables
\begin{equation}
u \equiv \frac{\gamma}{\beta}\,\rho \ \ \ \ \ \ \ \  \tau \equiv \frac{\beta^2}{\gamma}\,t \ \ \ \ \ \ \ \ \chi \equiv \frac{x}{R_c}
\end{equation} 
which give dimensionless version of Eq.\,\eqref{mf}:
\begin{equation}\label{non-dim}
   \dfrac{\partial u(\chi,\tau)}{\partial \tau} = \left[\overline{r}  +  \tilde{u}_{f}(\chi,\tau) -\tilde{u}_{c}^2(\chi,\tau)\right] u(\chi,\tau) + \overline{D} \,\nabla^2 _\chi u(\chi,\tau). 
\end{equation}
where
\begin{equation}
    \tilde{u}_f(\chi, \tau)=\int G\left(\left|\chi-\chi^{\prime}\right|,R\right) u\left(\chi^{\prime}, \tau\right) \mathrm{d} \chi^\prime
\end{equation}
and
\begin{equation}
    \tilde{u}_c(\chi, \tau)=\int G\left(\left|\chi-\chi^{\prime}\right|,1\right) u\left(\chi^{\prime}, \tau\right) \mathrm{d} \chi^\prime
\end{equation}
with scaled parameters 
\begin{equation}
    \overline{r} \equiv \frac{\gamma}{\beta ^2}\,r \ \ \ \ \ \  \overline{D}\equiv \frac{\gamma}{( R_c \,\beta)^2}\,D \ \ \ \ \ \ R\equiv \frac{R_f}{R_c}
\end{equation}

Using this dimensionless model description, the total, non-dimensional population abundance is given by
\begin{equation}
\mathcal{A}=\frac{\gamma}{\beta \, R_c} A.
\end{equation}

\section{Numerical methods}
\subsection{Stochastic individual-based simulations: the Gillespie algorithm}

We simulate the stochastic individual-based model using the Gillespie algorithm, which is a widely used and efficient method to generate realizations of a stochastic process \citep{gillespie1976general,gillespie1977exact}. The algorithm consists of two steps. First, we sample the time it takes for the next event to happen and, second, we sample which reaction takes place based on how each of them contributes to the total rate. The algorithm is based on the following steps: 

\begin{enumerate}
    \item At the beginning of the simulation, choose an initial condition for the number of individual in each lattice node.
    \item Following Section \ref{sec:AppMEQ}, compute all the possible global transition rates $\Omega(\eta \to \eta ')$ from the current configuration, $\eta$, to any other possible configuration $\eta'$. Define the total exit rate from the current configuration $\eta$,
    \begin{equation}
        \Omega_{\eta}^{\s{OUT}} = \displaystyle\sum_{\eta'}\Omega(\eta \to \eta').
    \end{equation}
    \item Sample the time to the next reaction, $\Delta t$, from an exponential distribution with mean equal to $1/\Omega_{\eta}^{\s{OUT}}$.
    \item Sample which of the possible transitions $\eta \to \eta'$ will take place. To do this sample, we define a probability of observing a transition to a specific configuration $\eta'$ as
    \begin{equation}
     P(\eta \to \eta')=\frac{\Omega(\eta \to \eta ')}{\Omega_{\eta}^{\s{OUT}}}
    \end{equation}
    \item Update the time and the configuration of the population to the new sampled values: $t\to t + \Delta t$ and $\eta \to \eta'$.
    \item Repeat steps 2 to 5 until the desired simulation time is reached.
\end{enumerate}

\subsection{Integration of the population-level approximation, Eq.\,(2.8)}

We perform the numerical integration of the population-level approximation PDE, Eq.\,(2.8) in the main text, using a second-order in time pseudospectral method detailed in \cite{montagne1997wound}. The method consists in determining the time evolution of a PDE in Fourier space through a ``two-step'' process after which we obtain the density field at a time $t+2\delta$ with an error $\mathcal{O}(\delta t^3)$. First, we Fourier transform the nonlinear PDE, Eq.\,(2.8) in the main text, and separate its linear and nonlinear terms
\begin{equation}
    \dfrac{\partial \hat{\rho}(k,t)}{\partial t} =  -\alpha(k) \hat{\rho}(k,t) + \Phi(k,t)
\end{equation}
where $\hat{\rho}(k,t)$ is the Fourier transform of the population density field and $\alpha(k)= D k^2 -r$ is the coefficient associated with the linear part. $\Phi(k,t)$ is the Fourier transform of the nonlinear part of the original equation
\begin{equation}
\Phi(k,t) = \mathcal{F}\bigg[ \tilde{\rho}_f(x,t) \rho(x,t)   - \gamma \, \tilde{\rho}_c^2(x,t) \rho(x,t) \bigg]
\end{equation}
 After setting the initial condition $\rho(x,0)$ and computing its Fourier transform $\hat{\rho}(k,0)$, the algorithm to integrate the equation between $t$ and $t+2\delta t$ is based on the following steps:

\begin{enumerate}
    \item Compute $\Phi(x,t)$ in real space and transform it to Fourier space to obtain $\Phi(k,t)$.
    
    \item Calculate the Fourier transform of the density field at time $t+\delta t$ (see \cite{montagne1997wound} for a derivation of this expression) as,
    \begin{equation}
    \displaystyle \hat{\rho}(k,t+\delta t)=e^{-\alpha(k) \delta t} \hat{\rho}(k,t)+\dfrac{1-e^{-\alpha(k) \delta t}}{\alpha(k)} \Phi(k,t)
    \end{equation}

    \item Compute $\rho(x,t+\delta t)$ by Fourier transforming $\hat{\rho}(k,t+\delta t)$ and use it to calculate the nonlinear part of the original PDE in real space. 
    
    \item Compute $\Phi(k,t+\delta t)$ by Fourier transforming the result obtained in the previous step.
    
    \item Calculate the updated field in the Fouier domain, 
    \begin{equation}
           \hat{\rho}(k,t+2\delta t)=e^{-2\alpha(k) \delta t} \hat{\rho}(k,t)+\dfrac{1-e^{-2\alpha(k) \delta t}}{\alpha(k)} \Phi(k,t+\delta t)
    \end{equation}
\end{enumerate}

Thus, in each algorithm iteration, the field $\hat{\rho}(k,t)$ goes to $\hat{\rho}(k,t+2\delta t)$ and the process is repeated until the desired simulation time is reached. For all of the population-level model simulations, we used $dt=0.05$, $dx=0.008$ and an initial condition $\rho_+ + \phi(x)$, where $\phi(x)$ is a noise uncorrelated in space with mean zero and variance $\epsilon\ll 1$. To compare PDE and IBM simulations, we transformed population densities to population sizes (abundance) by multiplying the value of the density field in each of the PDE integration nodes by the length of the lattice mesh used in the discrete simulations $\delta x$.

%\section{Supplementary Figure for the two-dimensional case}

%\begin{figure}[!h]\centering
%\includegraphics[width=\linewidth]{composition_2D.pdf}
%\caption{\label{fig:9} Model results in two dimensions. a) Spatial pattern of population density for $r=-1$. (b) Total abundance divided by the system size in the presence of aggregates. Results are obtained using the deterministic equation for population density (blue circles) and the group-level approximation (dashed red line). Filled and empty symbols represent the stable and unstable states, respectively. The deterministic simulations run until $\tau=3000$, $d\tau=0.05$. $\beta=1$, $\gamma=1$, $R_f=0.5$, $R_c=1$ and $ dx=0.008$, $L=10$ and $D=10^{-3}$.}
%\end{figure}

\bibliography{bibliography}